\documentclass[twocolumn]{article}
\let\ifapj\iffalse
\let\ifarxiv\iftrue
\let\iflocal\iffalse
\usepackage{ifpdf,ifxetex,ifluatex}
\ifxetex\else\ifluatex\else\usepackage[utf8]{inputenc}\fi\fi
\ifarxiv\usepackage{font-termes}\fi
\iflocal\usepackage{font-termes}\fi
\makeatletter
\@ifpackageloaded{unicode-math}{}{\let\symcal\mathcal}
\makeatother
\usepackage{etoolbox,ifthen}
\ifapj
  \usepackage[strict]{patch-apj}
  \usepackage{acro,csquotes,hyperref,siunitx}
\fi
\ifboolexpr{bool{arxiv} or bool{local}}{
  \usepackage{siunitx}
  \usepackage{basic,physics,astro}
  \usepackage[graytext]{draft}
  \usepackage[mode=biblatex, hyperref]{apjbib}
  \hypersetup{urlcolor=blue}
  \ExecuteBibliographyOptions{embedlinks}
  \addbibresource{tde.bib}
}{}
\ifluatex\hypersetup{pdfencoding=auto}\fi
\makeatletter
\ifpdf\@ifpackageloaded{astro}{\let\ast@symit\textit}{}\fi
\makeatother
\DeclareAcronym{TDE}{short=TDE, long=tidal disruption event}
\DeclareAcronym{AGN}{short=AGN, long=active galactic nucleus, short-indefinite=an,                long-indefinite=an, long-plural-form=active galactic nuclei}
\DeclareAcronym{MHD}{short=MHD, long=magnetohydrodynamic,     short-indefinite=an, short-plural=,                     long-plural=s}
\newcommand*\smc{\dot M_\su s}
\newcommand*\dmc{\dot M_\su d}
\newcommand*\amc{\dot M_\su a}
\newcommand*\mcr{\smc/\dmc}
\newcommand*\hrd{(H/R)_\su d}
\newcommand*\scmq{\biggl(\frac{M_\star}{M_\su h}\biggr)}
\newcommand*\scmmh{\biggl(\frac{M_\su h}{\SI{e6}{\solarmass}}\biggr)}
\newcommand*\scmms{\biggl(\frac{M_\star}{\si{\solarmass}}\biggr)}
\newcommand*\scmrs{\biggl(\frac{r_\star}{\si{\solarradius}}\biggr)}
\newcommand*\scmrp{\biggl(\frac{r_\su p}{r_\su t}\biggr)}
\newcommand*\scmrsrp
  {\biggl(\frac{r_\star}{\si{\solarradius}}\frac{r_\su p}{r_\su t}\biggr)}

\newcommand*\scmla{\biggl(\frac{L_\su a/L_\su E}{0.005}\biggr)}
\newcommand*\scmalpha{\biggl(\frac\alpha{0.1}\biggr)}
\newcommand*\scmeta{\biggl(\frac\eta{0.1}\biggr)}
\newcommand*\scmhrd{\biggl[\frac\hrd{\num{0.003}}\biggr]}
\newcommand*\sctmh{\allowbreak[M_\su h/(\SI{e6}{\solarmass})]}
\newcommand*\sctms{\allowbreak(M_\star/\si{\solarmass})}
\newcommand*\sctrs{\allowbreak(r_\star/\si{\solarradius})}
\newcommand*\sctrp{\allowbreak(r_\su p/r_\su t)}
\newcommand*\sctla{\allowbreak[(L_\su a/L_\su E)/0.005]}
\newcommand*\sctalpha{\allowbreak(\alpha/0.1)}
\newcommand*\scteta{\allowbreak(\eta/0.1)}
\newcommand*\scthrd{\allowbreak[\hrd/(\num{0.003})]}
\newcommand*\movie{(see \href{https://youtu.be/hF4tcCAFvA8}{movie})}
\ifapj\let\edit\relax\else\usepackage{xcolor}\fi
\expandafter\def\csname editcolor1\endcsname{blue!60!cyan}
\newcommand*\edit[2]{\textcolor{\csname editcolor#1\endcsname}{#2}}
\begin{document}

\title{Tidal disruption events in active galactic nuclei}

\ifapj
  \author[0000-0001-5949-6109]{Chi-Ho Chan}
  \affiliation{Racah Institute of Physics, Hebrew University of Jerusalem,
  Jerusalem 91904, Israel}
  \affiliation{School of Physics and Astronomy, Tel Aviv University,
  Tel Aviv 69978, Israel}
  \author[0000-0002-7964-5420]{Tsvi Piran}
  \affiliation{Racah Institute of Physics, Hebrew University of Jerusalem,
  Jerusalem 91904, Israel}
  \author[0000-0002-2995-7717]{Julian~H. Krolik}
  \affiliation{Department of Physics and Astronomy, Johns Hopkins University,
  Baltimore, MD 21218, USA}
  \author[0000-0002-0350-7419]{Dekel Saban}
  \affiliation{Racah Institute of Physics, Hebrew University of Jerusalem,
  Jerusalem 91904, Israel}
\fi

\ifboolexpr{bool{arxiv} or bool{local}}{
  \author[1,2]{Chi-Ho Chan}
  \author[1]{Tsvi Piran}
  \author[3]{Julian~H. Krolik}
  \author[1]{Dekel Saban}
  \affil[1]{Racah Institute of Physics, Hebrew University of Jerusalem,
  Jerusalem 91904, Israel}
  \affil[2]{School of Physics and Astronomy, Tel Aviv University,
  Tel Aviv 69978, Israel}
  \affil[3]{Department of Physics and Astronomy, Johns Hopkins University,
  Baltimore, MD 21218, USA}
}{}

\date{July 29, 2019}

\shorttitle{TDEs in AGNs}
\shortauthors{Chan et al.}
\pdftitle{Tidal disruption events in active galactic nuclei}
\pdfauthors{Chi-Ho Chan, Tsvi Piran, Julian Krolik, Dekel Saban}

\begin{abstract}
A fraction of \acp{TDE} occur in \acp{AGN} whose black holes possess accretion
disks; these \acp{TDE} can be confused with common \ac{AGN} flares. The
disruption itself is unaffected by the disk, but the evolution of the bound
debris stream is modified by its collision with the disk when it returns to
pericenter. The outcome of the collision is largely determined by the ratio of
the stream mass current to the azimuthal mass current of the disk rotating
underneath the stream footprint, which in turn depends on the mass and
luminosity of the \ac{AGN}. To characterize \acp{TDE} in \acp{AGN}, we
simulated a suite of stream--disk collisions with various mass current ratios.
The collision excites shocks in the disk, leading to inflow and energy
dissipation orders of magnitude above Eddington; however, much of the radiation
is trapped in the inflow and advected into the black hole, so the actual
bolometric luminosity may be closer to Eddington. The emergent spectrum may not
be thermal, \ac{TDE}-like, or \ac{AGN}-like. The rapid inflow causes the disk
interior to the impact point to be depleted within a fraction of the mass
return time. If the stream is heavy enough to penetrate the disk, part of the
outgoing material eventually hits the disk again, dissipating its kinetic
energy in the second collision; another part becomes unbound, emitting
synchrotron radiation as it shocks with surrounding gas.
\end{abstract}
\acresetall

\section{Introduction}

\Iac{TDE} occurs when a star wanders within the tidal radius of a black hole
and is ripped apart by tidal gravity; roughly half of the star is expelled as
unbound debris while the other half remains bound on highly eccentric orbits.
In the standard picture, general relativistic effects cause the bound debris
orbits to precess and self-intersect. Shocks at the intersections dissipate
kinetic energy, and the stream of bound debris is promptly gathered into an
accretion disk with radius approximately twice the stream pericenter distance
\citep[e.g.,][]{1988Natur.333..523R}. However, unless the stream pericenter is
within \num{\sim10} gravitational radii of the black hole, general relativistic
precession creates only weak, oblique shocks near apocenter that dissipate
energy inefficiently \citep{2015ApJ...804...85S, 2015ApJ...812L..39D}.
Consequently, the disk may not form right away, and most of the stream will
return to large distances. These weak shocks could be responsible for the
emission in optical \acp{TDE} \citep{2015ApJ...806..164P}.

This picture tacitly assumes that \acp{TDE} happen in vacuum. However, the
black hole in \iac{AGN} is surrounded by an accretion disk. As a star heads
toward the black hole on a trajectory destined for tidal disruption, its
initial passage through the disk leaves no lasting impact on either the star or
the disk because the density contrast between the two is immense. But after the
star is tidally disrupted, the bound debris stream has such a low density that,
when it returns to pericenter, it can interact with the disk in a more
interesting manner \citetext{\citealp{1994ApJ...422..508K}\multicitedelim
\bibstring{seealso}\space\citealp{2017MNRAS.469..314K}}. The collision between
the stream and the disk can potentially dissipate much of the kinetic energy
possessed by the stream and the disk gas near the impact point, which can be
radiated away. The collision can also alter the subsequent evolution of the
stream and drastically damage the disk. The collision with the disk likely
produces a brighter signal and has a greater effect on the stream than any
other interaction with circumnuclear material \citep{2016MNRAS.458.3324B}
because the disk is far denser.

Several percent of galaxies harbor \acp{AGN}, so a similar fraction of
\acp{TDE} should take place in \ac{AGN} hosts. This number is made uncertain to
the degree that the distribution of stellar orbits near \iac{AGN} is
systematically different from the center of an inactive galaxy, and that the
black hole in a galaxy with \iac{AGN} tends to be more massive than a galaxy
without one \citep[see also][]{2007A&A...470...11K, 2016MNRAS.460..240K}.
Because both \acp{TDE} and \acp{AGN} vary on timescales of weeks to months, and
because \iac{TDE} in \iac{AGN} presents less contrast against the prior state
of the system than \iac{TDE} in an inactive galaxy, deciding whether an
increase in brightness is due to \iac{TDE} or is merely \ac{AGN} variability is
not trivial \citep{2015JHEAp...7..148K, 2017NatAs...1..865K,
2018ApJ...852...37A, 2019NatAs...3..242T}; indeed, there are a number of cases
in which the correct identification of a particular episode of variation is
disputed \citep[e.g.,][]{2015A&A...581A..17C, 2015ApJ...803L..28G,
2015MNRAS.452...69M, 2015MNRAS.454.2798S, 2017ApJ...843..106B,
2017MNRAS.468..783L, 2017MNRAS.465L.114W, 2018Sci...361..482M,
2018ApJ...857L..16S}. It is therefore of interest to see if \acp{TDE} in
\acp{AGN} have distinctive observational characteristics that allow us to
recognize them more reliably.

Numerous physical processes interact in the course of these events. Radiation
is expected to contribute significantly to the internal energy in the parts of
the disk affected by the collision, while \acp{MHD} is critical to both radial
inflow and, at higher altitudes, vertical support. Consequently, a proper study
of the collision calls for simulations including both general relativistic
\acp{MHD} and the interaction between gas and radiation. Moreover, the
collision is described by a number of parameters: the black hole mass, the
stellar mass, the disk accretion rate, and the orientation of the stream with
respect to the disk. A large suite of simulations covering the realizable
subset of the multidimensional parameter space is needed to probe the full
range of observational behavior. As an exploratory step, here we present
pure-hydrodynamics simulations of one particular configuration of the
collision. Our simulations consider only the portion of the event when the mass
return rate is maximum, and they focus on the dependence on one parameter, the
ratio of the stream mass current to the azimuthal mass current of the disk
passing under the stream footprint. Nonetheless, even these simulations
identify a number of key mechanisms and reveal the principal issues that must
be resolved before making definite observational predictions.

We start off by estimating the properties of stream--disk collisions in
\cref{sec:analytics}. We introduce our simulation setup in \cref{sec:methods},
present our results in \cref{sec:results} \movie, and discuss possible
observational signatures in \cref{sec:discussion}. Our conclusions are
summarized in \cref{sec:conclusions}.

\section{Analytic considerations}
\label{sec:analytics}

A star of mass $M_\star$ and radius $r_\star$ is torn apart when it flies by a
black hole of mass $M_\su h$ on an orbit whose pericenter distance $r_\su p$ is
smaller than the tidal radius $r_\su t\eqdef r_\star(M_\star/M_\su h)^{-1/3}$
\citep{1975Natur.254..295H}. The tidal radius for a main-sequence star
interacting with a black hole commonly found in galactic nuclei is only several
tens of gravitational radii $r_\su g\eqdef GM_\su h/c^2$ of the black hole,
where $G$ is the gravitational constant and $c$ is the speed of light:
\begin{equation}\label{eq:tidal radius}
\frac{r_\su t}{r_\su g}\approx50\,\scmmh^{-2/3}\scmms^{-1/3}\scmrs.
\end{equation}
The portion of the debris remaining bound to the black hole traverses a highly
elliptical orbit and returns to pericenter as a stream. Different parts of the
stream have different semimajor axes; the part with semimajor axis $a$ returns
to pericenter at time $T=2\pi(GM_\su h/a^3)^{-1/2}$ after disruption. The most
bound part has the smallest semimajor axis $a_\su{mb}\sim\tfrac12r_\su
p^2/r_\star$ and returns to pericenter soonest, after a mass return time of
\begin{equation}\label{eq:mass return time}
T_\su{mb}\sim\frac\pi{\sqrt2}
  \biggl(\frac{GM_\su h}{r_\su p^3}\biggr)^{-1/2}
  \scmq^{-1/2}\scmrp^{3/2}.
\end{equation}
If the bound mass is uniformly distributed in specific binding energy $E=GM_\su
h/(2a)$, that is, $\ods{M_\star}E\sim\tfrac12M_\star/(GM_\su hr_\star/r_\su
p^2)$, the stream mass current, or the peak rate at which mass returns to
pericenter at early times, is
\begin{align}\label{eq:stream mass current}
\nonumber \smc &\sim
  \evalb*{\od{M_\star}E\,\abs*{\od Ea}\,\od aT}_{a=a_\su{mb}} \\
&\sim \frac{\sqrt2}{3\pi}M_\star
  \biggl(\frac{GM_\su h}{r_\su p^3}\biggr)^{1/2}
  \scmq^{1/2}\scmrp^{-3/2}.
\end{align}
Note that $\smc T_\su{mb}\sim\tfrac13M_\star$.

The black hole of \iac{AGN} has an accretion disk prior to the \ac{TDE}. The
stream intersects the disk at two points, the line joining which passes through
the black hole as well. Because stars originate from all directions, this line
is randomly oriented in the stream plane. Moreover, because the apocenter of a
highly eccentric orbit subtends only a small angle at the black hole, the
stream crosses the disk near pericenter much more often than near apocenter.
The configuration considered by \citet{2017MNRAS.469..314K}, in which the
stream travels within a thick, very weakly accreting disk and interaction
happens most strongly near apocenter, is special in comparison \citep[see
also][]{2017ApJ...843..106B}. For simplicity, we consider a parabolic stream
slamming perpendicularly into the disk at pericenter.

The stream interacts with the azimuthal mass current of the disk rotating
underneath its footprint, given by
\begin{equation}
\dmc\sim4R_\su s\Sigma_\su d
  \biggl(\frac{GM_\su h}{r_\su p}\biggr)^{1/2},
\end{equation}
where $R_\su s$ is the stream width at pericenter and $\Sigma_\su d$ is the
disk surface density at stream pericenter measured from the midplane to
infinity. A wider stream results in a larger $\dmc$, but $\smc$ is unchanged
because it is determined purely by orbital dynamics. The disk accretion rate,
\begin{equation}
\amc\sim2\pi r_\su p^2\Sigma_\su d\alpha\hrd^2
  \biggl(\frac{GM_\su h}{r_\su p^3}\biggr)^{1/2},
\end{equation}
with $\alpha$ the \citet{1973A&A....24..337S} parameter and $\hrd$ the disk
aspect ratio at stream pericenter, is much smaller than $\dmc$.

The ratio $\mcr$ is the most important quantity governing the outcome of the
collision, and it is the parameter we vary in our simulations. We would like to
relate $\mcr$ to the observables $M_\su h$ and $L_\su a$; here $L_\su
a=\eta\amc c^2$ is the disk accretion luminosity and $\eta$ is the accretion
efficiency. The value of $\dmc$ depends on the conditions of the disk. For
typical \acp{TDE} in \acp{AGN}, electron scattering dominates opacity in a
\citet{1973A&A....24..337S} disk at stream pericenter.
The nature of pressure switches from gas to radiation when
\begin{align}
\nonumber \frac{L_\su a}{L_\su E} &\gtrsim \num{0.004}\,
  \scmmh^{-1}\scmms^{-7/16}\contbin\times \\
&\cont \scmrsrp^{21/16}\scmalpha^{-1/8}\scmeta,
\end{align}
where $L_\su E=4\pi GM_\su hc/\kappaT$ is the Eddington luminosity and
$\kappaT$ is the cross section per mass for Thomson scattering. Now
$\dmc/\amc\propto\hrd^{-2}$, and $\hrd\propto\amc^{1/5}$ and $\hrd\propto\amc$
for gas- and radiation-dominated disks respectively; therefore, $\dmc$
increases with $L_\su a/L_\su E$ in a gas-dominated disk, then decreases with
$L_\su a/L_\su E$ as the disk becomes radiation-dominated.

\begin{figure}
\includegraphics{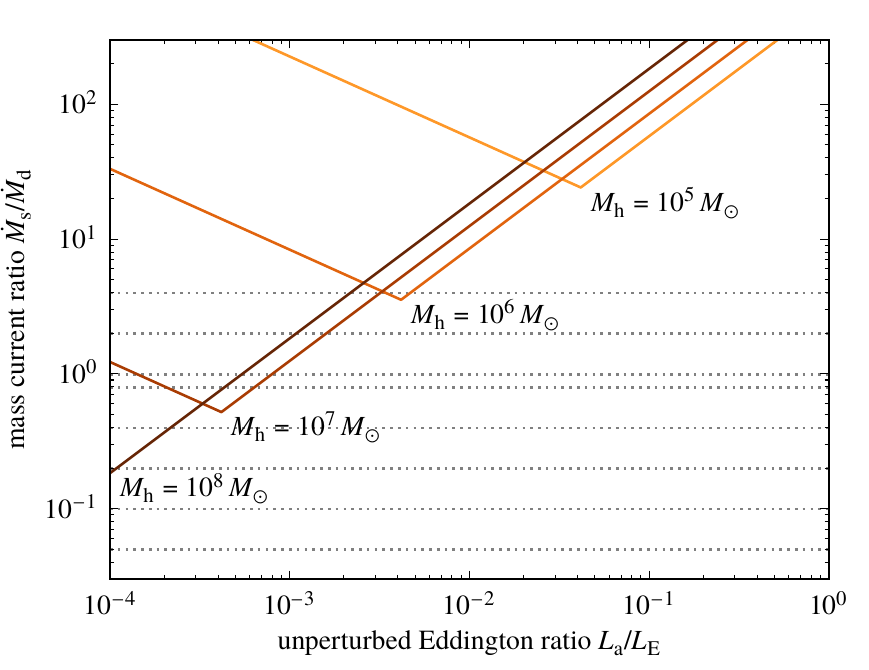}
\caption{Plot of the mass current ratio $\mcr$ as a function of the black hole
mass $M_\su h$ and the unperturbed disk Eddington ratio $L_\su a/L_\su E$ for
our fiducial parameters $M_\star=\si{\solarmass}$, $r_\star=\si{\solarradius}$,
$r_\su p=r_\su t$, $R_\su s=r_\star$, $\alpha=0.1$, and $\eta=0.1$
(\cref{sec:fiducial}). Solid lines reach a minimum when pressure in the disk
shifts from gas-dominated on the left to radiation-dominated on the right. The
horizontal lines are the $\mcr$ used in our simulations (\cref{sec:stream and
disk}).}
\label{fig:mass current ratio}
\end{figure}

\Cref{fig:mass current ratio} shows the dependence of $\mcr$ on $M_\su h$ and
$L_\su a/L_\su E$. Holding $M_\su h$ constant, a given level of $\mcr$ can be
realized in either a weakly accreting, gas-dominated disk or a strongly
accreting, radiation-dominated disk. The value of $\mcr$ tends to stay above
unity, dipping below only for the dimmest and most massive \acp{AGN}; this is
fortuitous because a number of our simulation results (\cref{fig:inflow
rate,fig:inflow time,fig:energy dissipation,fig:outgoing material
mass,fig:crude luminosity}) depend weakly on $\mcr$ once it is large enough.

We can estimate time-integrated consequences of the collision by comparing
global measures of stream and disk physical quantities. The mass ratio is
\begin{align}
\nonumber \frac{\tfrac12M_\star}{2\pi r_\su p^2\Sigma_\su d}
  &\sim 80\,
  \scmmh^{-1}\scmms^{3/2}\scmrsrp^{-3/2}\contbin\times \\
&\cont \scmla^{-1}\scmalpha\scmeta\scmhrd^2.
\end{align}
Because the stream velocity at pericenter is $\smash{\sqrt2}$ times the disk
orbital velocity there, the momentum ratio is of the same order as the mass
ratio. This means the stream carries enough mass and momentum to potentially
reshape and reorient the disk. Similarly, the ratio of stream kinetic energy to
disk binding energy is of order the mass ratio; therefore, if all the stream
kinetic energy were dissipated, the stream could heat the disk sufficiently to
unbind it.

If the dissipated energy were instead radiated away immediately, the collision
could be quite luminous:
\begin{equation}
\frac{\smc(GM_\su h/r_\su p)}{L_\su E}
  \sim 30\,\scmmh^{-5/6}\scmms^{7/3}\scmrs^{-5/2}\scmrp^{-4},
\end{equation}
corresponding to $\smc(GM_\su h/r_\su p)/L_\su a\approx6000$ for a disk with
$L_\su a/L_\su E\approx0.005$. We shall discuss in \cref{sec:radiative
transfer} why the actual luminosity could be very different.

\section{Methods}
\label{sec:methods}

The stream--disk collision is simulated with Athena++, a rewrite of the
finite-volume hydrodynamics code Athena \citep{2008ApJS..178..137S}. We adopt
the cylindrical coordinates $(R,\phi,z)$. Due to the coordinate singularity at
$R=0$, we must restrict the lower-radial boundary of our simulation domain to
$R>0$, thus introducing a cylindrical cutout in the center of our simulation
domain. Gas directed toward the cutout is removed from the simulation entirely.
Other details of our numerical setup follow.

\subsection{Equations}

The stream pericenter distance of typical \acp{TDE} is only tens of $r_\su g$
(\cref{sec:analytics}). Given other approximations used, it is reasonable to
neglect relativistic effects. The hydrodynamics equations are
\begin{align}
\label{eq:mass}
\pd\rho t+\divg(\rho\vec v) &= 0, \\
\label{eq:momentum}
\pd{}t(\rho\vec v)+\divg(\rho\vec v\vec v+p\tsr I) &= -\rho\grad\Phi, \\
\label{eq:energy}
\pd Et+\divg[(E+p)\vec v] &= -\rho\vec v\cdot\grad\Phi.
\end{align}
Here $\rho$, $\vec v$, and $p$ are density, velocity, and pressure,
$\Phi(R,z)=-GM_\su h/(R^2+z^2)^{1/2}$ is the gravitational potential of the
black hole, and $\tsr I$ is the isotropic rank-two tensor. We use an adiabatic
equation of state. The total energy is $E=\tfrac12\rho v^2+\rho e$, where
$\rho e=p/(\gamma-1)$ is the internal energy. Disk pressure can be dominated
by gas or radiation (\cref{sec:analytics}); for simplicity, we choose the
adiabatic index to be $\gamma=\tfrac53$. As we shall see later in
\cref{sec:thermodynamics}, the thermodynamic conditions of a realistic disk is
likely more complicated than a simple choice between gas and radiation
pressure.

\subsection{Code units}
\label{sec:code units}

Simulation quantities are expressed in code units of length $r_0$, time
$\Omega_0^{-1}\eqdef(GM_\su h/r_0^3)^{-1/2}$, velocity $v_0\eqdef r_0\Omega_0$,
and density $\rho_0$. Because Newtonian gravity, unlike relativistic gravity,
is scale-free, the dimensionless versions of
\cref{eq:mass,eq:momentum,eq:energy} in this unit system are independent of
$r_0$ and $\rho_0$. With an appropriate choice of these two quantities, our
results can be scaled to the conditions of any particular \ac{TDE}.

We set $r_0=r_\su p$, the characteristic length scale of the system. Time is
reported as the number of disk orbits at $R=r_0$, or disk orbits for short; one
disk orbit is $2\pi\sctrp^{3/2}$ times the stellar dynamical time. The mass
return time in \cref{eq:mass return time} is
\begin{equation}\label{eq:mass return time conversion}
T_\su{mb}\sim400\,
  \scmmh^{1/2}\scmms^{-1/2}\scmrp^{3/2}\,\mtext{disk orbits}.
\end{equation}

The value of $\rho_0$ will be determined in \cref{sec:density}.

\subsection{Fiducial parameters}
\label{sec:fiducial}

When translating simulation results from code units to physical units, we adopt
the same fiducial parameters as in \cref{sec:analytics}, to wit, $M_\su
h=\SI{e6}{\solarmass}$, $M_\star=\si{\solarmass}$, $r_\star=\si{\solarradius}$,
$r_\su p=r_\su t$, $R_\su s=r_\star$, $\alpha=0.1$, and $\eta=0.1$.

\subsection{Stream and disk properties}
\label{sec:stream and disk}

The initial disk has a constant midplane density $\rho_0$ and a Gaussian scale
height $\mathrelp\propto R$:
\begin{equation}
\rho(R,z)=\rho_0\exp[-\Delta\Phi/(\symcal H^2v_\su K^2)].
\end{equation}
Here $\Delta\Phi(R,z)\eqdef\Phi(R,z)-\Phi(R,0)$ is the gravitational potential
difference from the midplane and is $\mathrelp\approx\tfrac12v_\su K^2(z/R)^2$
for our spherically symmetric $\Phi$, where $v_\su K^2(R)\eqdef(\pds\Phi{\ln
R})_{z=0}$ is the square of the midplane Keplerian orbital velocity. The aspect
ratio of the simulated disk is $\symcal H=0.1$; it is much larger than the
aspect ratio of a \citet{1973A&A....24..337S} disk for numerical reasons, and
the effect of this choice will be considered in \cref{sec:density}. The
pressure of the disk is $p(R,z)=\rho\symcal H^2v_\su K^2$. Its orbital
velocity, given by
\begin{equation}
v_\phi^2(R,z)=
  v_\su K^2+(\Delta\Phi+\symcal H^2v_\su K^2)(\pds{\ln v_\su K^2}{\ln R}),
\end{equation}
is slightly sub-Keplerian to counteract pressure forces.

Our initial disk is not a \citet{1973A&A....24..337S} disk. For example, its
surface density profile is
\begin{equation}\label{eq:surface density}
\Sigma(R)=\int_0^{z_\su{max}}dz\,
  \rho\approx(\tfrac12\pi)^{1/2}\symcal H\rho_0R,
\end{equation}
where $z_\su{max}$ is the distance of the vertical boundaries of the simulation
domain from the midplane. This surface density profile is neither the
$\Sigma\propto R^{-3/5}$ profile of a gas-dominated disk nor the $\Sigma\propto
R^{3/2}$ profile of a radiation-dominated disk, although it is close to the
latter. The advantage of our initial disk is that it is scale-free, so we can
scale our results to fit any \ac{TDE} of interest. As long as we pick $\rho_0$
such that $\mcr$ is the same for our initial disk and for a
\citet{1973A&A....24..337S} disk, our results are not qualitatively affected.

Ideally, we would like our stream to travel on a parabolic trajectory vertical
to the midplane: The stream would approach pericenter from $\phi=\pi$, cross
the axis, reach pericenter at $(R,\phi,z)=(r_0,0,0)$, and return to infinity
along $\phi=\pi$. Unfortunately, the trajectory would then cross the
cylindrical cutout in the center of the simulation domain. We solve this
problem in two ways. For lighter streams that are effectively stopped by the
disk, we can simply lower the upper-vertical boundary of the simulation domain
until the stream intersects the boundary at $\phi=0$; if the stream is injected
as a boundary condition from the intersection, it will not encounter the cutout
before it terminates at pericenter. For heavier streams that can punch through
the disk, we additionally rotate the trajectory \SI{0.15}{\radian} from the
vertical to make it avoid the cutout altogether. The sense of the rotation is
to make the trajectory prograde with respect to the disk, increasing the
likelihood that the outgoing material will miss the cutout. Since the vertical
and inclined streams differ so little in inclination, we treat them as directly
comparable. We leave the exploration of streams of other spatial orientations
to future work.

We choose the stream boundary condition so that, if the stream traveled
ballistically to pericenter, its cross section there would be circular and its
transverse density profile would be a Gaussian of radius $\symcal Wr_\su 0$,
where $\symcal W=0.02$. The simulated stream width is slightly wider than in
typical \acp{TDE}. The density at the center of the Gaussian is set by matching
the desired value of $\smc$, and the pressure is $\num{e-6}\,v_0^2$ times
density. Tracing orbits back to the upper-vertical boundary determines the
stream density, velocity, and pressure there.

The disk mass current is $\dmc\approx0.01\,\rho_0r_0^2v_0$. We use
$\smc\in\{0.5,1,2,4,8\}\times\num{e-3}\,\rho_0r_0^2v_0$ for the vertical stream
and $\smc\in\{1,2,4\}\times\num{e-2}\,\rho_0r_0^2v_0$ for the inclined stream.
Altogether, we have $\mcr\in\{%
\num{\approx0.05},\allowbreak\num{\approx0.1},\allowbreak
\num{\approx0.2},\allowbreak\num{\approx0.4},\allowbreak
\num{\approx0.8},\allowbreak\num{\approx1},\allowbreak
\num{\approx2},\allowbreak\num{\approx4}\}$. Combinations of $M_\su h$ and
$L_\su a$ giving such $\mcr$ can be read off from \cref{fig:mass current
ratio}. Only the heaviest stream pertains to our fiducial parameters
(\cref{sec:fiducial}); the lighter streams are also valid if $M_\su h$ is
larger or $\alpha$ is smaller.

\label{sec:stream threshold}

The initial disk and the stream have respectively $\abs j\lesssim0.2\,r_0v_0$
and $j\gtrsim1.4\,r_0v_0$, where $j$ is the specific angular momentum in the
$\phi=\tfrac12\pi$ direction. For the purpose of separating disk-like gas from
stream-like gas when analyzing our results in \cref{sec:results}, we adopt the
conservative cutoff $j=1.1\,r_0v_0$.

\subsection{Stream and disk density}
\label{sec:density}

The aspect ratio of a \citet{1973A&A....24..337S} disk at stream pericenter,
$\hrd\sim\num{0.003}$, and the stream width, $R_\su s\sim r_\star$, are both
difficult to resolve spatially; this is why the simulated disk height and
stream width are artificially increased to $\symcal Hr_0$ and $\symcal Wr_0$
respectively, with $\symcal H=0.1$ and $\symcal W=0.02$ (\cref{sec:stream and
disk}). To preserve the mass currents under such a modification, we
simultaneously adjust the midplane density of the disk and the central density
of the stream. We demand that the simulated disk mass current, $4\symcal
Wr_0\Sigma(r_0)v_0$, be equal to $\dmc$ of a realistic disk. This, together
with \cref{eq:stream mass current,eq:surface density}, yields
\begin{equation}\label{eq:density unit}
\rho_0=\frac1{6\pi^{3/2}\symcal H\symcal W}\frac{M_\star}{r_\star^3}
  \scmq^{3/2}\scmrp^{-9/2}(\mcr)^{-1}.
\end{equation}
Similarly, the central density of the stream is determined by matching the
desired $\smc$. Because the mass currents involved in the collision are
independent of the particular values of $\symcal H$ and $\symcal W$ provided
both are $\mathrelp\ll1$, the outcome of the collision should not be severely
affected by our choices for these quantities. Volume-integrated quantities in
our simulations, such as the inflow rate (\cref{sec:inner disk mass}) and the
energy dissipation rate (\cref{sec:energy dissipation}), are likewise genuine.
Although the simulated disk is unrealistically thick at the beginning, any
structure created in the course of the simulations with aspect ratio
$\mathrelp\gg\symcal H$ is realistic because its thickness is due to the
injection of internal energy much greater than the artificial internal energy
in the initial condition.

\subsection{Other numerical considerations}

Periodic boundary conditions are used for the azimuthal direction. Outflow
boundary conditions are used for the radial and vertical directions: Velocity
is copied from the last physical cell into the ghost zone and inward-pointing
velocity components are zeroed, while density and pressure are isothermally
extrapolated such that the pressure gradient balances gravitational and
centrifugal forces in the ghost zone. On the upper-vertical boundary, the
stream injection condition (\cref{sec:stream and disk}) supersedes this
boundary condition wherever the former predicts a larger density.

The numerical vacuum is an axisymmetric hydrostatic torus. Its density is
determined by the constraint
\begin{equation}\label{eq:vacuum density}
\const=-\Phi(R,z)+\frac{(v_\phi^2)_\su v}{2-2q}
  -K\frac\Gamma{\Gamma-1}\rho_\su v^{\Gamma-1},
\end{equation}
its velocity is $\vec v_\su v=v_0(R/r_0)^{1-q}\,\uvec e_\phi$, and its pressure
is $p_\su v=K\rho_\su v^\Gamma$. The parameters in these equations are
$q=1.75$, $K=0.5\,\rho_0^{1-\Gamma}v_0^2$, and $\Gamma=0.9$; the constant on
the left-hand side of \cref{eq:vacuum density} follows from requiring that
$\rho_\su v$ attain a maximum of $10^{-6}\rho_0$ at $(R,z)=(r_0,0)$. As the
simulation progresses, we keep the density and pressure of every cell greater
than or equal to their vacuum values at all times. In addition, whenever the
density of a cell drops to $0<\xi<1$ times vacuum, we simultaneously modify its
velocity to $\xi\vec v+(1-\xi)\vec v_\su v$ so as to prevent velocities from
erroneously growing in low-density regions.

The simulation domain spans
$[0.15\,r_0,5\,r_0]\times[-\pi,\pi]\times[-3.2\,r_0,1.6\,r_0]$ in $(R,\phi,z)$.
The inner-radial boundary at $R=0.15\,r_0$ encloses a cylindrical cutout in the
center of the simulation domain, introduced to exclude the coordinate
singularity at $R=0$. The radius of the cutout in physical units is
$\num{\approx7}\,r_\su g\, \sctmh^{-2/3}\sctms^{-1/3}\sctrs\sctrp$, which is
just outside the innermost stable circular orbit of a nonrotating black hole;
therefore, we are simulating the entire disk interior to the stream pericenter,
and we may regard gas entering the cutout as falling into the black hole. The
lower-vertical boundary is twice as far below the midplane as the
upper-vertical boundary so that we can follow as much of the gas leaving the
lower side of the disk as feasible.

The number of grid cells is $200\times200\times300$. We employ a power-law grid
in the radial direction for which $\Delta R_{i+1}/\Delta R_i=1.01$; this policy
gives us cells with approximately square poloidal cross sections at $R=r_0$.
The pressure scale height of the disk at stream pericenter,
$\mathrelp\approx\smash{\sqrt2}\symcal Hr_0$, is resolved with \num{\approx9}
vertical cells, and the stream width, $\symcal Wr_0$, is barely resolved with
\num{\approx2} azimuthal cells at the injection point.

The simulations run to 100 disk orbits, which, according to \cref{eq:mass
return time conversion}, is
$\num{\sim0.3}\,T_\su{mb}\,\sctmh^{-1/2}\sctms^{1/2}\sctrp^{-3/2}$. This means
our simulations focus on the time when the mass return rate is greatest, and
the mass return rate varies little over the course of our simulations.

\section{Results}
\label{sec:results}

\subsection{Overview}
\label{sec:overview}

\begin{figure*}
\includegraphics{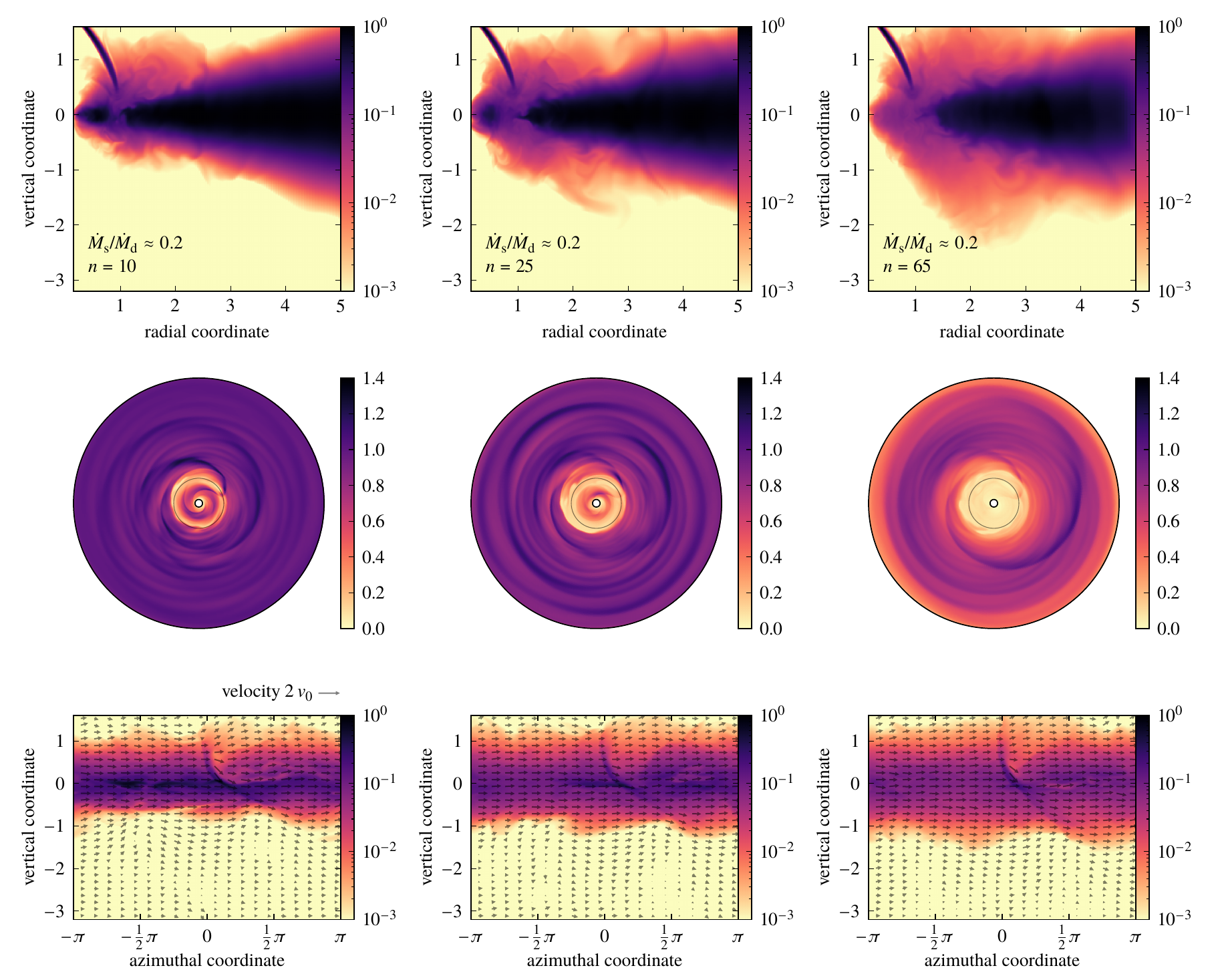}
\caption{Orthogonal slices of the $\mcr\approx0.2$ simulation \movie. Each
column presents one snapshot with the number $n$ of disk orbits as indicated in
the top row. The top, middle, and bottom rows are respectively a poloidal slice
at $\phi=0$, a midplane slice, and a cylindrical slice at $R=r_0$. The slices
intersect at $(R,\phi,z)=(r_0,0,0)$; it is where the stream would reach
pericenter if the disk were absent, and approximately where the stream collides
with the disk when the disk is present. Colors plot density; the color scale is
logarithmic in the top and bottom rows and linear in the middle row. The gray
circle in the middle row marks $R=r_0$. Arrows in the bottom row show velocity;
the arrow above the left panel has length $2\,v_0$.}
\label{fig:evolution}
\end{figure*}

We use the $\mcr\approx0.2$ simulation to illustrate how the disk in any
simulation evolves in general \movie. The left column of \cref{fig:evolution}
displays an early time, 10 disk orbits. The incoming stream is visible above
the disk in the poloidal slice. The vertical structure near the center of the
cylindrical slice is the part of the stream closest to pericenter; it is bent
because the disk deflects the stream in the direction of disk rotation. The
stream opens an annular gap in the disk at $R\approx r_0$, manifest in the
poloidal and midplane slices; the gap separates the inner disk (\cref{sec:inner
disk}) at $R\le r_0$ from the outer disk.

Nonaxisymmetric features appear in the midplane slice. The most salient one is
the curved bow shock with its tip at $(R,\phi)\approx(r_0,0.1\pi)$; here
$\phi>0$ because of stream deflection. The inner half of the bow shock extends
inward, forming a prominent spiral shock in the inner disk. There are typically
multiple spiral shocks; together, they drive an extremely powerful inflow that
is orders of magnitude stronger than in the unperturbed disk (\cref{sec:inner
disk dynamics}).

Nonaxisymmetric features in the outer disk can be understood with the help of
the cylindrical slice. The incoming stream pushes stream and disk gas out of
the other side of the disk. Part of this gas reaches as far as $z\approx-r_0$
before gravity pulls it back down to strike the disk at
$\phi\approx-\tfrac34\pi$; the impact compresses the disk and launches a spiral
shock stretching outward from $(R,\phi)\approx(1.2\,r_0,0.8\pi)$. The impacting
gas glances off the disk and falls back to it once again at
$\phi\approx\tfrac38\pi$, launching another, much weaker, spiral feature. At
the time shown, the weak spiral feature is hidden from view by the similarly
located and much stronger bow shock, but it becomes more conspicuous at late
times when the bow shock is weaker. Both spiral features are stationary in
space.

The center column depicts an intermediate time, 25 disk orbits. The gap widens
and the spiral shocks deplete the inner disk, as evidenced by the poloidal and
midplane slices. The reduction of disk gas at $R\lesssim1.2\,r_0$ is the reason
why the stream suffers less deflection in the cylindrical slice. The collision
heats the inner disk, causing its gas to puff up and move to larger radii,
easily seen by comparing the poloidal slices of early and intermediate times.

The intermediate time is taken during an episode of disk evolution in which the
outer edge of the gap, as witnessed in the midplane slice, becomes highly
nonaxisymmetric; as the outer edge orbits around, acoustic waves are sent
propagating outward at the same frequency as its orbital frequency. These waves
have large enough amplitudes to obscure the spiral features in the outer disk.

The right column presents a late time, 65 disk orbits. The inner disk is
largely cleared out; as a result, spiral shocks in the inner disk are barely
visible. The outer edge of the cavity returns to approximate axisymmetry, so
waves are no longer launched and the two spiral features in the outer disk
re-emerge. The spiral shock extending outward from
$(R,\phi)\approx(1.5\,r_0,0.8\pi)$ remains well-defined, but the weak spiral
feature starting from $(R,\phi)\approx(2.6\,r_0,0.4\pi)$ is merely a diffuse
density enhancement. The outer half of the bow shock appears as a spur joining
the weak spiral feature.

\begin{figure}
\includegraphics{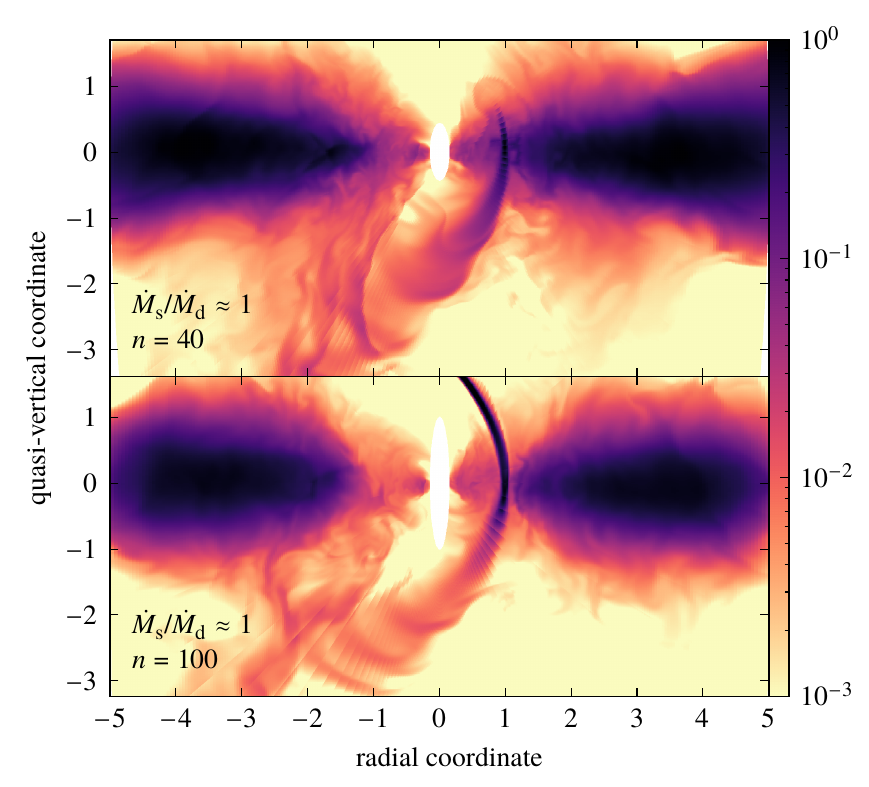}
\caption{Inclined slices of the $\mcr\approx1$ simulation. Each panel presents
one snapshot with the number $n$ of disk orbits as indicated in the bottom-left
corner. Both panels are slices passing through the origin and the stream
pericenter, but the top and bottom panels are rotated \SI{0.35}{\radian} and
\SI{0.15}{\radian} respectively from the vertical, to match the orbital plane
of the outgoing material. The stream is more strongly deflected at early times,
so only at late times can we pick a slice where both the incoming stream and
the outgoing material are simultaneously visible. Colors plot density.}
\label{fig:outgoing material}
\end{figure}

The incoming stream in the $\mcr\approx0.2$ simulation is not heavy enough to
produce a perceptible stream of outgoing material (\cref{sec:outgoing
material}); for a better view of the outgoing material, we turn to the
$\mcr\approx1$ simulation. The slices shown in \cref{fig:outgoing material}
roughly follow the orbital plane of the outgoing material. The outgoing
material exits the disk not as a dense, collimated structure, but as a clumpy
plume orders of magnitude more dilute than the incoming stream. The plume is
more spread out in the orbital plane than perpendicular to it.

\subsection{Inner disk}
\label{sec:inner disk}

\subsubsection{Dynamics}
\label{sec:inner disk dynamics}

\begin{figure*}
\includegraphics{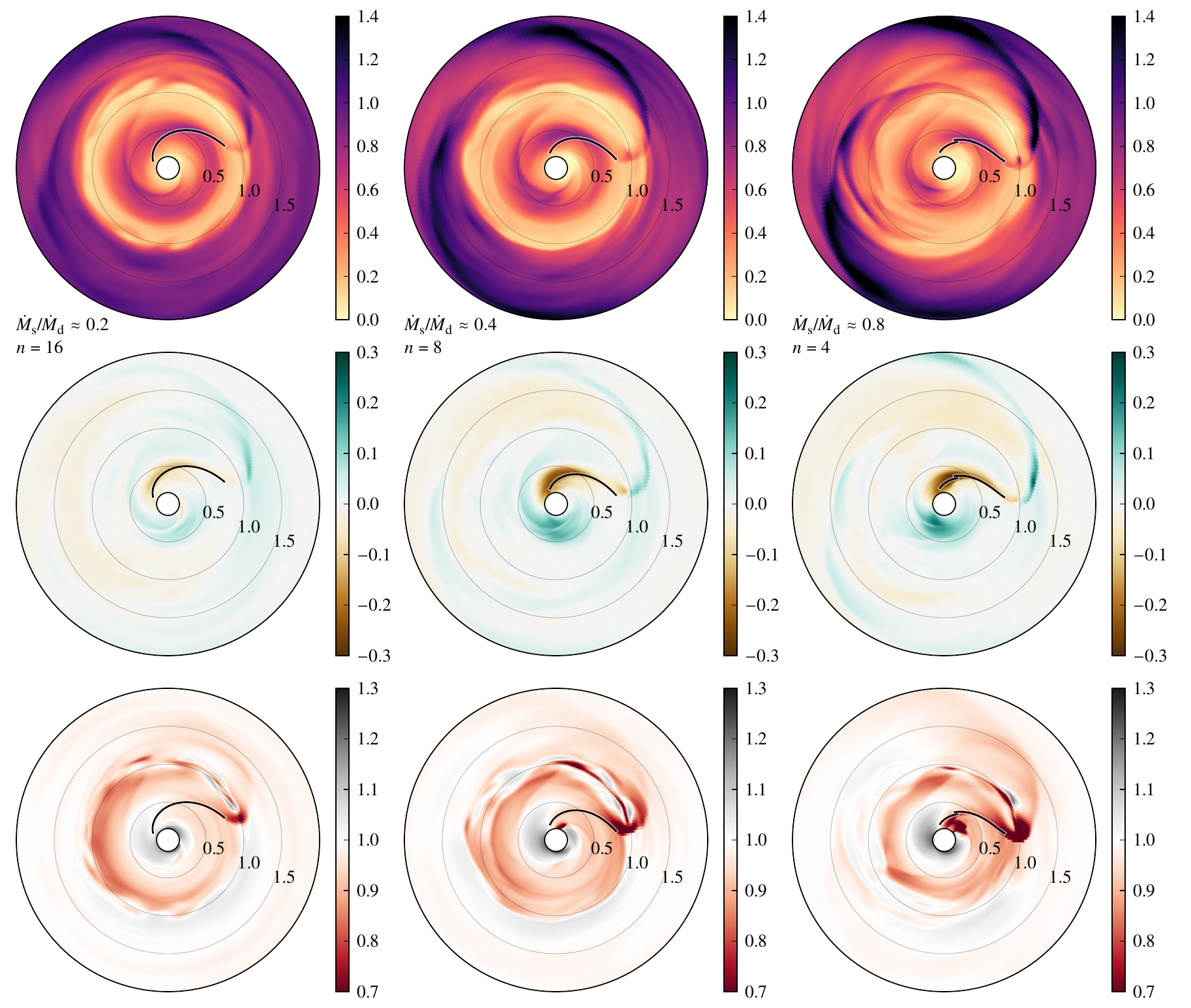}
\caption{Zoomed-in midplane slices. Each column presents one snapshot with mass
current ratio $\mcr$ and the number $n$ of disk orbits as indicated between the
top and middle rows. Colors in the top row plot density, in the middle row the
radial mass flux, and in the bottom row the azimuthal velocity divided by the
Keplerian orbital velocity. The black arcs above the cutout trace where the
radial mass flux, the quantity plotted in the middle row, is the most strongly
inward.}
\label{fig:inner disk}
\end{figure*}

\Cref{fig:inner disk} displays three snapshots from three simulations. The
snapshots are selected from early times when the inner disk is still largely
intact, but our comments below hold for all simulations at all times. At the
times shown, the stream has delivered the same amount of mass, momentum, and
energy across all simulations.

The top row demonstrates how the collision excites spiral shocks in the inner
disk. Spiral shocks vary in number, position, and strength over time, but there
is often a dominant pair, one connected to the bow shock, the other located
almost directly opposite in the other half of the inner disk. The former spiral
shock is typically stronger and is marked with black arcs in the figure.

Spiral shocks deflect orbiting gas. The stronger spiral shock tends to deflect
gas inward, giving rise to a region of inward mass flux immediately after the
shock in the middle row. The net effect of the multiple spiral shocks is to
remove angular momentum from the orbiting gas; as a result, gas falls toward
the black hole on gradually shrinking, tightly wound trajectories. As is
evident in the bottom row, rotation departs more and more from Keplerian as gas
moves inward. Gas ultimately plunges into the cutout, most of it doing so over
only a small fraction of the circumference, and then into the black hole.

\subsubsection{Mass}
\label{sec:inner disk mass}

\begin{figure}
\includegraphics{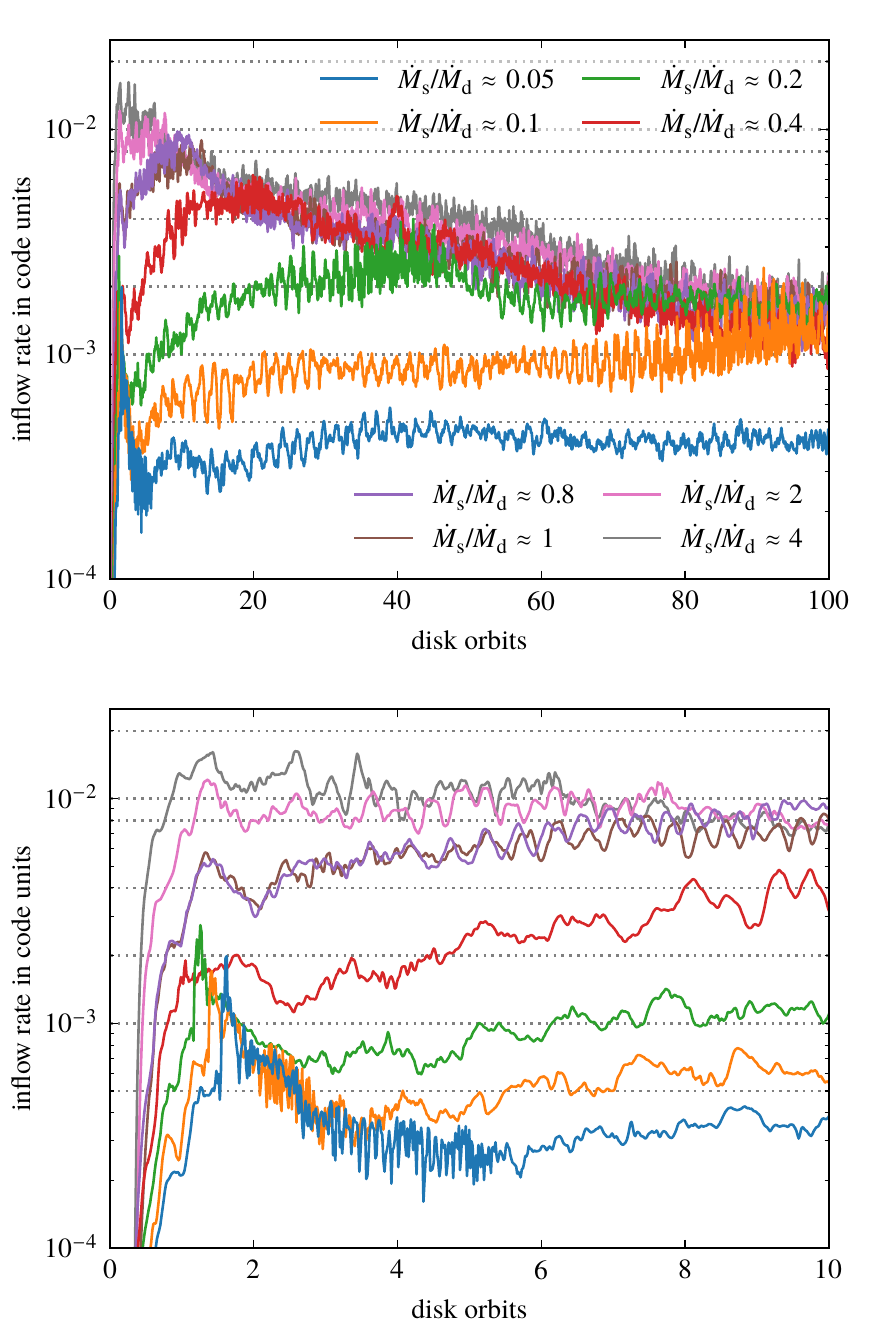}
\caption{Plot of the disk-like (\cref{sec:stream threshold}) inflow rate across
the inner-radial boundary as a function of time. The code unit is
$\rho_0r_0^2v_0$, where $\rho_0$ is from \cref{eq:density unit}, and $r_0$ and
$v_0$ are from \cref{sec:code units}. The top and bottom panels differ only in
that the top panel looks at long-term behavior while the bottom panel zooms in
on early times. The horizontal lines mark the stream mass current $\smc$ in
ascending order from bottom to top, omitting the largest $\smc$.}
\label{fig:inflow rate}
\end{figure}

The spiral shocks are extremely efficient at destroying the inner disk. For a
radiation-dominated, $\mcr\approx4$ disk with our fiducial parameters
(\cref{sec:fiducial}), the inflow rate across the inner-radial boundary given
by \cref{fig:inflow rate} is \numrange{\sim1.2e3}{1.2e4} times the unperturbed
level, or \numrange{\sim6}{50} times Eddington! In effect, the coherent spiral
shocks exert an extremely strong stress on the flow, leading to an inflow rate
much faster than ordinary disk processes, such as correlated \ac{MHD}
turbulence, could cause.

For $\mcr\lesssim0.2$, the inflow rate has a spike at early times lasting a few
disk orbits, then relaxes at late times to $\mathrelp\sim\smc$. For
$\mcr\gtrsim0.4$, the inflow rate rises in the first \numrange{\sim1}{20} disk
orbits, then decays gradually as the inner disk is depleted
(\cref{sec:overview}); the inflow rate at \num{\gtrsim50} disk orbits is nearly
independent of $\mcr$ because the stream appears essentially as a solid
obstacle to the disk, deflecting a fixed fraction of the inner disk toward the
black hole every disk orbit.

For $\mcr\lesssim1$, the inflow rate is quasiperiodically modulated over at
least part of the simulation. The quasiperiodic variation is particularly
strong for $0.1\lesssim\mcr\lesssim0.2$ and is visible in the top panel
throughout the simulation. This quasiperiodicity is due to the interaction
between the stationary spiral shocks and some orbiting nonaxisymmetric feature,
such as the lopsided outer edge of the gap (\cref{sec:overview}).

\begin{figure}
\includegraphics{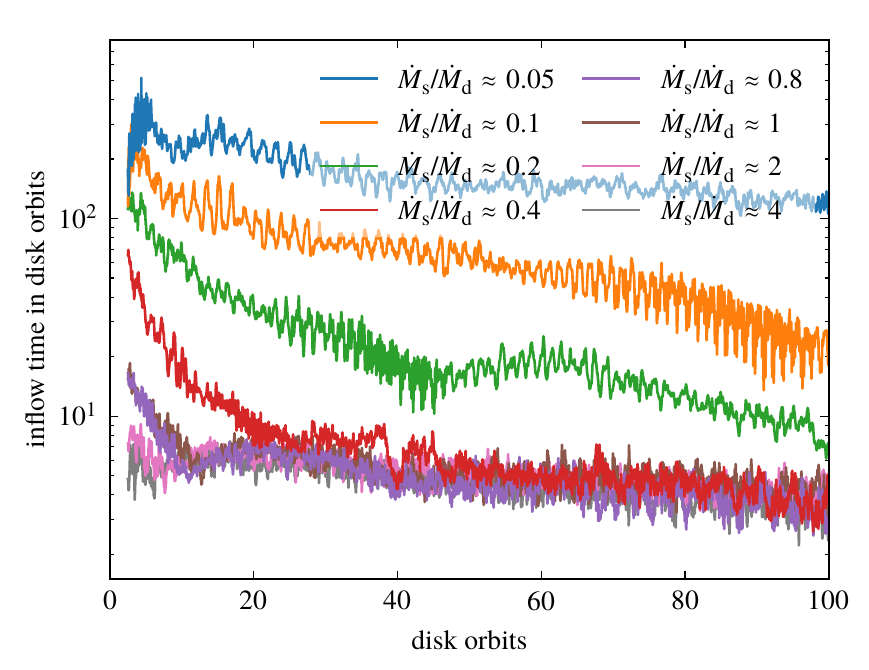}
\caption{Plot of the inflow time, defined as the disk-like (\cref{sec:stream
threshold}) mass at $R\le r_0$ divided by the disk-like inflow rate across the
inner-radial boundary (\cref{fig:inflow rate}), as a function of time.}
\label{fig:inflow time}
\end{figure}

\Cref{fig:inflow time} shows the inflow time, or the time it takes a gas packet
in the inner disk to move into the cutout on its way to the black hole. Thanks
to the spiral shocks, the inflow time is orders of magnitude shorter than that
of an unperturbed \citet{1973A&A....24..337S} disk measured at stream
pericenter, which is $\num{\sim1.8e5}\,\sctalpha^{-1}\scthrd^{-2}$ disk orbits,
but the multistep nature of the shock-driven inflow mechanism (\cref{sec:inner
disk dynamics}) means the inflow time is still at least a few disk orbits.

\begin{figure}
\includegraphics{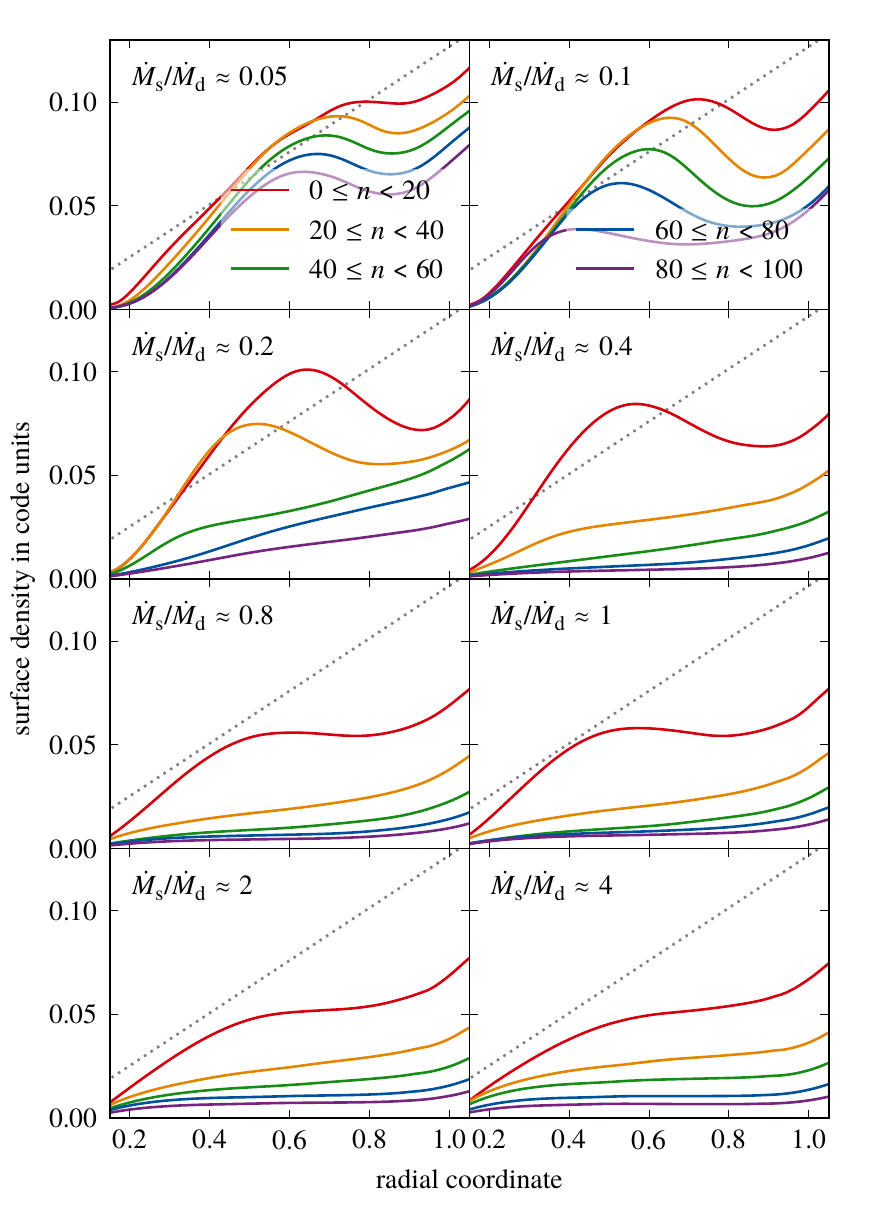}
\caption{Plot of the azimuthally averaged disk-like (\cref{sec:stream
threshold}) surface density, defined as half of the integral of density from
one side of the inner disk to another, as a function of radius. The code unit
is $\rho_0r_0$, where $\rho_0$ is from \cref{eq:density unit} and $r_0$ is from
\cref{sec:code units}. The dotted line shows the unperturbed disk surface
density given by \cref{eq:surface density}; the colored curves are
time-averages over the number $n$ of disk orbits specified in the legend. Each
panel presents one simulation with mass current ratio $\mcr$ as indicated in
the top-left corner.}
\label{fig:surface density}
\end{figure}

The inflow time is not the characteristic timescale on which the inner disk
mass decreases; this is because, as we shall see in \cref{sec:outgoing
material}, the inner disk captures enough of the stream to resupply itself.
\Cref{fig:surface density} demonstrates this considerably slower depletion of
the inner disk. Although the surface density declines monotonically with time,
by the end of the simulations at 100 disk orbits, which is
$\num{\sim0.3}\,T_\su{mb}$ for our fiducial parameters (\cref{sec:fiducial}),
the surface density is only about an order of magnitude lower than its
unperturbed value. Because a radiation-dominated, $\mcr\approx4$ disk with our
fiducial parameters (\cref{sec:fiducial}) starts out with a Thomson optical
depth of \num{\sim8000}, it stays optically thick to electron scattering
throughout the simulations.

\subsection{Energy dissipation}
\label{sec:energy dissipation}

\begin{figure}
\includegraphics{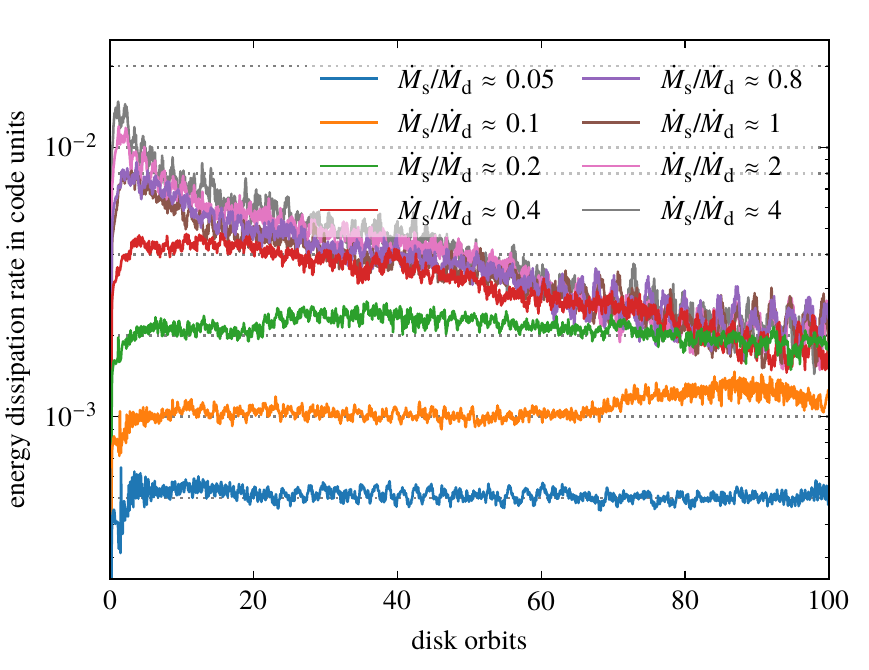}
\caption{Plot of the energy dissipation rate, defined as the volume integral of
$\pds{(\rho e)}t+\divg(\rho e\vec v)+p\divg\vec v$ over the simulation domain,
as a function of time. The code unit is $\rho_0r_0^2v_0^3$, where $\rho_0$ is
from \cref{eq:density unit}, and $r_0$ and $v_0$ are from \cref{sec:code
units}. The horizontal lines mark the stream kinetic energy current for
different stream mass current $\smc$ in ascending order from bottom to top,
omitting the largest $\smc$.}
\label{fig:energy dissipation}
\end{figure}

\Cref{fig:energy dissipation} tracks the time evolution of the energy
dissipation rate. For a radiation-dominated, $\mcr\approx4$ disk with our
fiducial parameters (\cref{sec:fiducial}), the energy dissipation rate is
\numrange{\sim300}{2000} times the unperturbed disk accretion luminosity, or
\numrange{\sim1.5}{10} times the Eddington luminosity.

For $\mcr\lesssim0.2$, the energy dissipation rate is always close to the
stream kinetic energy current; therefore, energy dissipation is well described
as due to a perfectly inelastic collision between the stream and an immovable
disk. For $\mcr\gtrsim0.8$, the energy dissipation rate initially rises but
eventually falls; the energy dissipation rate shows little dependence on $\mcr$
at \num{\gtrsim15} disk orbits. Because the disk cannot fully stop the stream
in these simulations, less of the stream kinetic energy is dissipated. Instead,
the heavy stream acts as a stationary barrier to orbiting disk gas, and disk
kinetic energy is dissipated when disk gas runs into the stream or the shocks
created by the collision. The falloff in the energy dissipation rate over time
is the result of the disk density decreasing (\cref{sec:overview}). For
$\mcr\approx0.4$ lying between the two extremes, the behavior of the energy
dissipation rate switches from resembling lighter-stream simulations in the
beginning to resembling heavier-stream simulations later on, suggesting that
the stream takes time to clear out a gap in the disk before it can go through
with little impediment.

\subsection{Outgoing material}
\label{sec:outgoing material}

\begin{figure}
\includegraphics{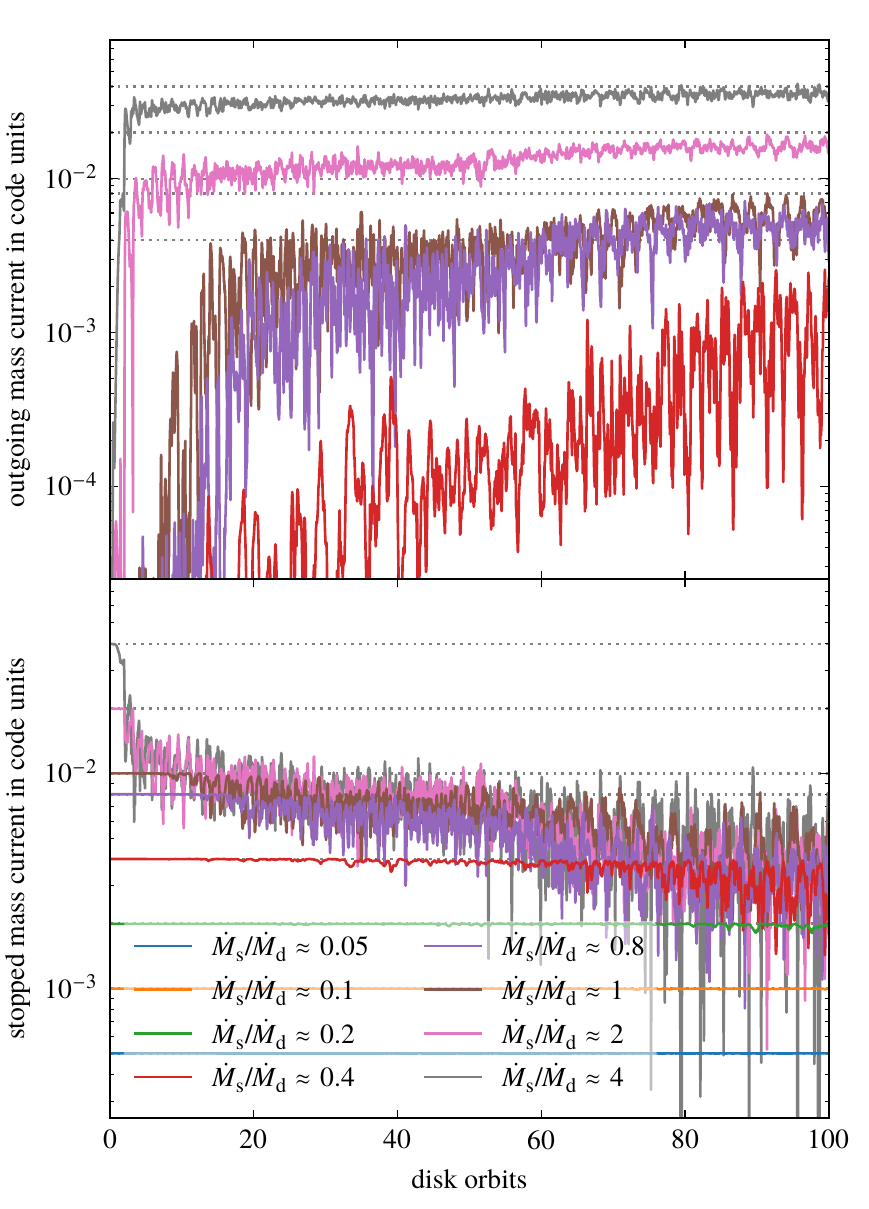}
\caption{\textit{Top panel:} Plot of the mass current of the stream-like
(\cref{sec:stream threshold}) outgoing material emerging from the bottom of the
disk as a function of time. The code unit is $\rho_0r_0^2v_0$, where $\rho_0$
is from \cref{eq:density unit}, and $r_0$ and $v_0$ are from \cref{sec:code
units}. Simulations with negligible outgoing material are hidden. The
horizontal lines mark the stream mass current $\smc$ for the simulations shown.
\textit{Bottom panel:} Plot of the stopped mass current, defined as $\smc$
minus the outgoing material mass current depicted in the top panel. The
horizontal lines mark $\smc$ for each simulation. The stopped mass current is
generally larger than the inflow rate in \cref{fig:inflow rate}, implying that
the stream carries enough mass to potentially resupply the inner disk.}
\label{fig:outgoing material mass}
\end{figure}

The top panel of \cref{fig:outgoing material mass} plots the outgoing material
mass current, showing only the simulations for which the outgoing material has
a mass current \SI{\ge1}{\percent} that of the incoming stream throughout most
of the simulation. A greater fraction of the stream passes through at later
times because the disk density is lower (\cref{sec:overview}), but there are
large and rapid fluctuations about the overall rising trend.

The bottom panel displays the mass current of the incoming stream stopped by
the disk. For $\mcr\lesssim0.2$, the stopped mass current is
$\mathrelp\approx\smc$, which in turn is approximately the inflow rate in
\cref{fig:inflow rate}. We may therefore picture the stream as being absorbed
into the disk and deflected straight toward the black hole; the latter part is
consistent with the fact that the disk removes all kinetic energy from lighter
streams (\cref{sec:energy dissipation}). For $\mcr\gtrsim0.8$, the disk shaves
off only a fraction of the incoming stream because the inner disk is cleared
out early on (\cref{sec:inner disk mass}).

In all cases, the stopped mass current is almost always larger than the inflow
rate in \cref{fig:inflow rate}. Mass loss from the inner disk (\cref{sec:inner
disk mass}) slows down significantly if just a fraction of this mass current is
diverted to resupply the inner disk. The inner disk mass still decreases over
time because the outer disk receives part of the stopped mass current; because
gas splashes back from the upper side of the disk; and because the collision
heats up the inner disk (\cref{sec:overview}), causing it to expand out through
the upper-vertical boundary or outward in radius.

Since the outgoing material is rather cold when it reaches the lower-vertical
boundary, with sound speed $\mathrelp\lesssim0.1\,v_0$, its dynamics beyond the
simulation domain is dictated by its mechanical energy. \Cref{fig:outgoing
material mechanical energy} shows the specific mechanical energy distribution
of the outgoing material, again only for simulations with substantial outgoing
material. The most bound part has specific binding energy
$\num{\sim0.05}\,GM_\su h/r_0$, hence it flies out on an elliptical orbit to
\num{\sim19} times the pericenter distance, then returns to pericenter after
$\num{\sim0.09}\,T_\su{mb}\,\sctmh^{-1/2}\sctms^{1/2}\sctrp^{-3/2}$.

\begin{figure}
\includegraphics{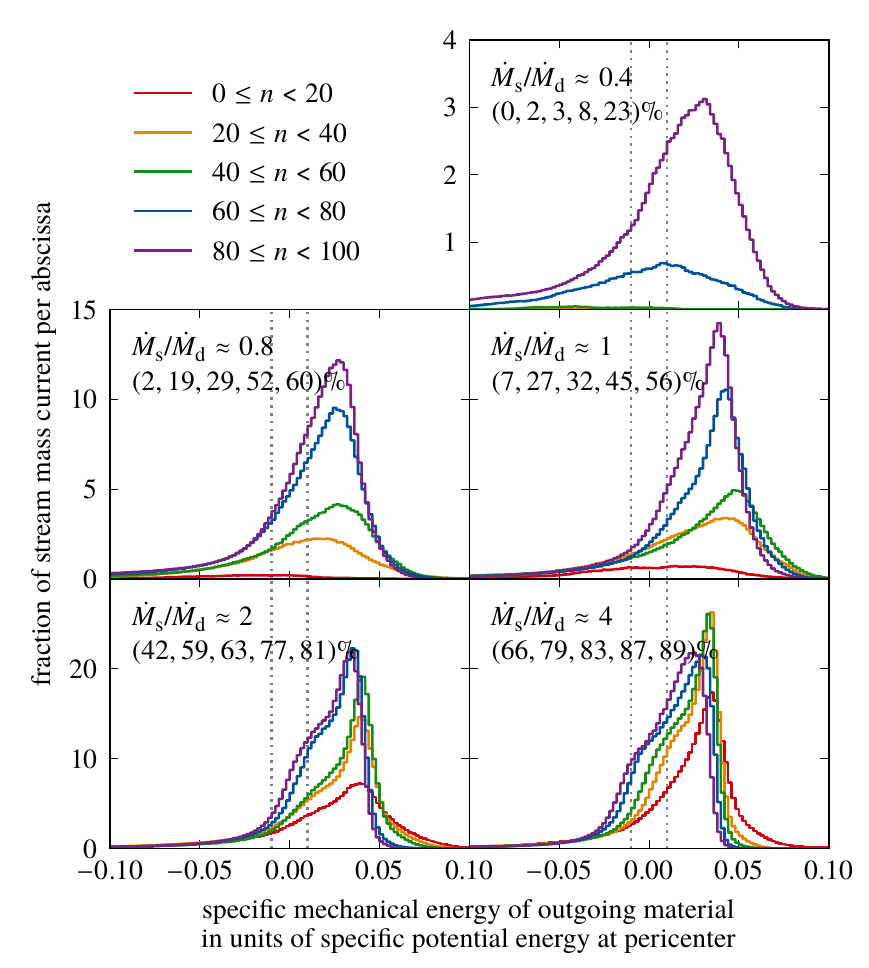}
\caption{Histogram of the mass current--weighted specific mechanical energy of
stream-like (\cref{sec:stream threshold}) outgoing material. Each panel
presents one simulation with mass current ratio $\mcr$ as indicated in the
top-left corner; only simulations with the largest $\mcr$ are included. The
time intervals over which the histograms are computed are given in the legend
in terms of the number $n$ of disk orbits. Each histogram is normalized such
that the area under it is the fraction of the incoming stream ending up in the
outgoing material; these areas are given in the top-left corner underneath
$\mcr$. Note that the ordinate range varies from row to row. The vertical lines
indicate the characteristic energy spread the initial disruption imparts to the
stellar material for our fiducial parameters $M_\su h=\SI{e6}{\solarmass}$,
$M_\star=\si{\solarmass}$, and $r_\star=\si{\solarradius}$
(\cref{sec:fiducial}).}
\label{fig:outgoing material mechanical energy}
\end{figure}

Most of the outgoing material, having received a kick from the disk during the
collision, is unbound with specific mechanical energy $\num{\sim0.03}\,GM_\su
h/r_0$. The outgoing material carries much less energy than is dissipated in
the collision (\cref{sec:energy dissipation}), but is much more energetic than
the debris unbound during the initial disruption. The disk is less dense at
later times (\cref{sec:overview}) and imparts a smaller force on the stream,
thus the outgoing material is less and less unbound after \num{\sim40} disk
orbits.

Two caveats must be noted. First, we use a marginally bound parabolic incoming
stream in the simulations, whereas a realistic incoming stream has specific
binding energy $\num{\sim0.01}\,GM_\su h/r_0\times
\sctmh^{-1/3}\sctms^{1/3}\sctrp^{-1}$. We therefore expect the specific
mechanical energy distribution of realistic outgoing material to be shifted
toward lower energies by a similar amount, changing the fraction of bound and
unbound material. Second, the bound outgoing material may not return to
pericenter if it is intercepted on the way by the large-scale disk
\citep{2017MNRAS.469..314K}.

\section{Discussion}
\label{sec:discussion}

\subsection{Timescales}

\Acp{TDE} in \acp{AGN} are governed by several timescales. Here we group them
together in ways that allow meaningful comparisons.

\subsubsection{Timescales of unperturbed disk flows}
\label{sec:unperturbed timescales}

The shortest and longest timescales of the system both pertain to the
unperturbed disk. The disk dynamical time at stream pericenter is $1/(2\pi)$
disk orbit or
$\SI{\sim0.018}{\day}\times\sctms^{-1/2}\sctrs^{3/2}\sctrp^{3/2}$. It is
$\sctrp^{3/2}$ times the stellar dynamical time. Because outgoing waves
(\cref{sec:overview}) are launched into the outer disk by the time-varying
conditions near the impact point, the frequency of these waves is approximately
the inverse of the disk dynamical time.

The unperturbed accretion time of a \citet{1973A&A....24..337S} disk at stream
pericenter is $\SI{\sim60}{\year}\times
\sctms^{-1/2}\sctrs^{3/2}\sctrp^{3/2}\sctalpha^{-1}\scthrd^{-2}$.

\subsubsection{Timescales of flows caused by the
\texorpdfstring{\ac{TDE}}{TDE}}
\label{sec:perturbed timescales}

The mass return time is given by \cref{eq:mass return time conversion}; in
physical units, it is
$\SI{\sim40}{\day}\times\sctmh^{1/2}\sctms^{-1}\sctrs^{3/2}\sctrp^3$. It is the
timescale on which the incoming stream, and hence the energy dissipation rate
in the collision, taper off. Because this timescale is sufficiently longer than
the duration of our simulations, we can approximate the mass return rate to be
constant.

The inflow time is the time a gas packet takes to fall toward the black hole,
hence also the time over which the rest energy of the gas packet is converted
to internal energy. For heavier streams, the inflow time declines slowly over
time from
\SIrange{\sim1}{0.5}{\day}${}\times\sctms^{-1/2}\sctrs^{3/2}\sctrp^{3/2}$; for
the lightest stream, it decreases from
\SIrange{\sim50}{12}{\day}${}\times\sctms^{-1/2}\sctrs^{3/2}\sctrp^{3/2}$
(\cref{sec:inner disk mass}).

If the stream is heavy enough to drill through the disk, part of the outgoing
material will be bound. The bound material travels on an elliptical orbit of
period $\SI{\sim4}{\day}\times\sctms^{-1/2}\sctrs^{3/2}\sctrp^{3/2}$
(\cref{sec:outgoing material}), which is considerably shorter than the mass
return time. The bound material will encounter the disk a second time, but
because it is much more dilute than the incoming stream (\cref{sec:overview}),
this time it will simply adhere to the disk and not go through; a rise in the
energy dissipation rate accompanies this interaction. As the incoming stream
weakens over time and less outgoing material is ejected, this extra energy
dissipation ceases as well.

\label{sec:magnetic stresses}

The collision leaves the disk in a very perturbed state. The disk returns to
its unperturbed state when the depleted inner disk (\cref{sec:overview}) is
refilled with gas from the outer disk. The timescale is very uncertain because
our simulations do not take into account how magnetic stresses pull gas in. For
a \citet{1973A&A....24..337S} disk, the resupply time is simply the unperturbed
accretion time (\cref{sec:unperturbed timescales}). However, if gas pressure
falls sharply from the outer disk to the inner, magnetic loops can stretch
inward across the interface. Orbital shear then creates magnetic stresses
strong enough to pull gas along with the loops, allowing gas to spiral inward
over \num{\sim10} disk orbits, or
$\SI{\sim1.2}{\day}\times\sctms^{-1/2}\sctrs^{3/2}\sctrp^{3/2}$. Such effects
are seen generically in the initial transient phase of global \ac{MHD} disk
simulations \citep{2011ApJ...743..115N}. Because the disk already captures
enough mass from the stream to greatly reduce its mass loss rate
(\cref{sec:outgoing material}), even a small amount of resupply could allow the
mass and surface density of the inner disk to reach steady state within a
fraction of the mass return time.

\subsubsection{Timescales of energy release}
\label{sec:energy release timescales}

The cooling time is the time it takes radiation originating from the midplane
to diffuse out of the geometrically (\cref{sec:overview}) and optically
(\cref{sec:inner disk mass}) thick inner disk; it is $\tau_\su TH/c$, where $H$
is the height of the perturbed inner disk measured at stream pericenter. For an
inner disk of aspect ratio unity, the cooling time is
\begin{equation}
\SI{\sim30}{\day}\times\scmmh^{1/3}\scmms^{-1/3}\scmrsrp
  \biggl(\frac{\tau_\su T}{\num{e4}}\biggr).
\end{equation}
Here $\tau_\su T$ is the Thomson optical depth at stream pericenter
(\cref{fig:surface density}). For a radiation-dominated, $\mcr\approx4$ disk
with our fiducial parameters (\cref{sec:fiducial}), the cooling time drops from
\num{\sim21} to \SI{\sim1.4}{\day} over the course of \SI{\approx12}{\day}. The
cooling time is so long because $\tau_\su T$ is \num{\sim0.1} times the
unperturbed Thomson optical depth (\cref{sec:inner disk mass}) but $H$ rises
greatly due to sudden heating. How quickly $\tau_\su T$ declines depends on the
amount of resupply (\cref{sec:magnetic stresses}).

Because the inner disk is optically thick to scattering, the inflow time
(\cref{sec:perturbed timescales}) is also the time radiation has to escape
before being swallowed. The ratio of the inflow time to the cooling time is a
rough estimate of the fraction of internal energy generated in the collision
that is released as radiation. For a radiation-dominated, $\mcr\approx4$ disk
with our fiducial parameters (\cref{sec:fiducial}), the ratio rises from
\num{\sim0.06} to \num{\sim0.3} over \SI{\approx12}{\day}. Most of the energy
is therefore advected into the black hole, suppressing the collision
luminosity. The situation may change if the inner disk is further depleted and
its optical depth diminished accordingly.

\subsubsection{Hierarchy}

For a radiation-dominated, $\mcr\approx4$ disk with our fiducial parameters
(\cref{sec:fiducial}), the shortest timescale is the disk dynamical time,
\SI{\sim0.018}{\day}. Next in order is the inflow time due to shocks, bottoming
out at \SIrange{\sim0.5}{1}{\day} within \SI{\sim1}{\day} from the beginning of
the collision. Magnetic stresses could bring gas from the outer disk into the
inner disk over \SI{\sim1.2}{\day}, while the bound outgoing material falls
back to the disk in \SI{\sim4}{\day}. These timescales are followed by the
cooling time due to radiative diffusion from the inner disk, which starts at
\SI{\sim21}{\day} but drops to \SI{\sim1.4}{\day} within \SI{\approx12}{\day}.
Longer still is the mass return time, \SI{\sim40}{\day}. The unperturbed disk
accretion time, \SI{\sim60}{\year}, is much longer than all of the other
timescales and is therefore irrelevant.

\subsection{Energetics}
\label{sec:energetics}

We present an inventory of the time-integrated energy we may expect over the
course of the entire event, including the energy from processes not directly
simulated. All items except the last are concerned with internal energy
production in the disk; the energy emitted as light from the disk could be much
smaller because radiation could be trapped in the inflow and swallowed by the
black hole (\cref{sec:energy release timescales}).


For lighter streams, the stream loses all of its kinetic energy at the
collision (\cref{sec:energy dissipation}); its gas assimilates into the disk
(\cref{sec:outgoing material}) and moves inward to the black hole within an
inflow time. The energy dissipated in this whole process is the same as if the
stream were accreted directly onto the black hole, that is,
$E_\star\sim\SI{9e52}{\erg}\times\sctms\scteta$. The energy dissipated may be
smaller because shocks in the inner disk may send gas straight into the black
hole (\cref{sec:inner disk dynamics}).

In the meantime, shocks excited by the collision dump the pre-existing inner
disk onto the black hole (\cref{sec:overview}). The energy produced is $E_\su
d\sim\SI{6e50}{\erg}\times
\sctmh^{-1/3}\sctms^{-7/6}\sctrs^{7/2}\sctrp^{7/2}\sctalpha^{-1}\scteta^2
\sctla^{-1}$ for a radiation-dominated inner disk. It can be smaller if shocked
gas plunges into the black hole (\cref{sec:inner disk dynamics}) or if only
part of the pre-existing inner disk is accreted.

After the event is over, magnetic stresses gradually replenish the inner disk
with gas from the outer disk (\cref{sec:energy release timescales}). Energy of
order the binding energy of the unperturbed inner disk is released, which is
$E_\su r\sim\SI{9e49}{\erg}\times
\sctmh^{1/3}\sctms^{-5/6}\sctrs^{5/2}\sctrp^{5/2}\sctalpha^{-1}\scteta
\sctla^{-1}$ if the inner disk is radiation-dominated. Should magnetic stresses
be strong enough to keep the inner disk refilled even while the event is in
progress (\cref{sec:magnetic stresses}), the continuous inflow of gas from the
outer disk to the black hole during the event may lead to a total amount of
energy dissipated exceeding $E_\su d+E_\su r$.

For heavier streams, the situation is more complicated. The inner disk is
similarly flushed out by shocks, generating $E_\su d$ of internal energy in the
process, and the inner disk is likewise resupplied by the outer disk, yielding
$E_\su r$. The difference is that the disk now captures only a fraction $f_\su
c\sim0.5$ of the incoming stream (\cref{sec:outgoing material}), and the rest
of the stream emerges on the other side as outgoing material. The energy
available from the initial dissipation at the impact point and from the
subsequent inflow of stream gas is therefore only $\mathrelp\approx f_\su
cE_\star$. The value of $f_\su c$ decreases with $\mcr$ and time, the decrease
being the sharpest around $\mcr\sim1$ (\cref{sec:outgoing material}).

Some of the outgoing material remains bound, with a range of specific binding
energy from $\mathrelp\sim0.05\,GM_\su h/r_\su p$ to zero (\cref{sec:outgoing
material}). Different parts of the bound material fly out to their respective
apocentric distances of $\mathrelp\gtrsim19\,r_\su p$, at which point they run
into the disk again. Because the bound material is much more diffuse than the
incoming stream was at pericenter (\cref{sec:overview}), we expect the disk to
capture all of the bound material. The kinetic energy dissipated is
$\SI{\lesssim2e49}{\erg}\times \sctms^2\sctrs^{-1}\sctrp^{-3}\allowbreak(f_\su
b/0.5)$, with $f_\su b\le1-f_\su c$ the fraction of the incoming stream ending
up in the bound material; therefore, the second collision may be too dim to be
seen against the first collision and the background \ac{AGN}. The captured
bound material accretes along with the disk, but the accretion energy can be
neglected because the accretion timescale is much longer than the mass return
time.

The previous paragraph pertains to the particular stream configuration used in
our simulations, one where the stream plane is perpendicular to the disk plane
and the stream pericenter is in the disk plane (\cref{sec:stream and disk}). In
the general case where the two planes are oriented randomly, it is more likely
that the first and second collisions both occur near pericenter. Such a second
collision dissipates
$\SI{\sim5e51}{\erg}\times\sctmh^{2/3}\sctms^{4/3}\sctrs^{-1}
\sctrp^{-1}\allowbreak(f_\su b/0.25)$, much greater than if the second
collision were at apocenter. This energy is a sizable fraction of that obtained
by accreting the bound material directly, $f_\su bE_\star$, because $r_\su
t/r_\su g$ is typically only a few times $\eta^{-1}$, according to
\cref{eq:tidal radius}. The dissipation occurs over large areas of the disk and
possibly at high altitudes, both of which have implications for how much of the
dissipated energy would emerge as radiation. The remainder of $f_\su bE_\star$
is liberated when the bound material falls toward the black hole.

Some of the outgoing material is unbound by a kick from the disk and carries
specific kinetic energy $\mathrelp\sim0.03\,GM_\su h/r_\su p$ at infinity
(\cref{sec:outgoing material}). The total energy is
$\SI{\sim1.4e50}{\erg}\times\sctmh^{2/3}\sctms^{4/3}\sctrs^{-1}
\sctrp^{-1}\allowbreak(f_\su u/0.25)$, where $f_\su u=1-f_\su c-f_\su b$ is the
fraction of the incoming stream unbound by the collision. Compared to the
primary ejecta unbound as an immediate consequence of the disruption
\citep{2016ApJ...822...48G, 2016ApJ...827..127K, 2019MNRAS.487.4083Y}, this
secondary ejecta has a fraction $\mathrelp\approx f_\su u$ of the mass, higher
velocity (\cref{sec:outgoing material}), and a larger opening angle
(\cref{sec:overview}). The ejecta drives a bow shock while running into the
surrounding medium, and relativistic electrons accelerated in this shock can
produce synchrotron radiation \citep{2016ApJ...827..127K, 2019MNRAS.487.4083Y}.
At later times, the shock driven by the ejecta may mimic a supernova remnant
\citep{2016ApJ...822...48G}. The denser medium around \acp{AGN} could mean that
\acp{TDE} are more radio-bright in \acp{AGN} than in vacuum. The prompt
emission due to the secondary ejecta could be more luminous than the primary
ejecta because the fastest material has higher velocity and a wider interaction
area \citep{2016ApJ...827..127K, 2019MNRAS.487.4083Y}. All these may explain
why the radio transient Cygnus~A\nobreakdash-2 \citep{2017ApJ...841..117P}, if
interpreted as a thermal \ac{TDE} happening in \iac{AGN}
\citep{2019MNRAS.486.3388D}, is brighter in radio than typical thermal
\acp{TDE} in vacuum.

\subsection{Inner disk}

\subsubsection{Thermodynamics}
\label{sec:thermodynamics}

Our simulations do not correctly track temperatures because the adiabatic index
used corresponds to gas pressure and not radiation pressure. To estimate the
gas and radiation temperatures of the inner disk, we observe that shocks raise
the sound speed of the simulated inner disk to \numrange{\sim0.2}{0.5} times
the local Keplerian orbital velocity; this corresponds to a gas temperature of
\begin{equation}
\SI{\sim2e10}{\kelvin}\times
  \scmmh^{2/3}\scmms^{1/3}
  \biggl(\frac{r_\star}{\si{\solarradius}}
  \frac R{r_\su t}\biggr)^{-1}.
\end{equation}
If gas and radiation in the inner disk had enough time to thermalize, they
would come into an equilibrium temperature of
\begin{align}
\nonumber T_\su{eq}\sim\SI{6e5}{\kelvin}\times
  \scmmh^{1/6}\scmms^{1/12}\contbin\times \\
\cont\biggl(\frac{r_\star}{\si{\solarradius}}
  \frac R{r_\su t}\biggr)^{-1/4}
  \biggl(\frac\rho{\SI{2e-10}{\gram\per\centi\meter\cubed}}\biggr)^{1/4},
\end{align}
where $\rho$ is the inner disk density during the collision, and its fiducial
value is the typical density at the end of the $\mcr\approx4$ simulation for a
radiation-dominated disk with our fiducial parameters (\cref{sec:fiducial}). In
thermodynamic equilibrium, the internal energy is dominated by radiation and
varies $\mathrelp\propto T_\su{eq}^4$, hence $T_\su{eq}$ is rather insensitive
to the specific values of the parameters on the right-hand side.

However, it may be difficult for gas raised to $T\sim\SI{e10}{\kelvin}$ to
reach thermodynamic equilibrium. Because the cross section per mass for
free--free absorption is $\mathrelp\propto\rho T^{-7/2}$, gas this hot would
have essentially zero absorptivity and therefore, by Kirchhoff's law,
essentially zero emissivity. The post-shock gas could cool by inverse Compton
scattering off photons already present in the gas, but this, too, is
problematic. The cooling rate is proportional to the local radiation energy
density, which depends strongly on whether the gas originates from the colder
stream or the hotter unperturbed inner disk. Even if Compton cooling is rapid,
it creates no new photons, so the temperature decrease it achieves may be
limited. Substantial Compton cooling may be possible only at high latitudes,
where the post-shock gas is exposed to photons radiated by the outer disk.
Close to the midplane, Compton cooling may create photons energetic enough to
produce pairs, which could hasten thermalization. For all these reasons,
without detailed calculations, we cannot make a firm statement about how
rapidly, or through what radiation mechanisms, the inner disk may approach
thermodynamic equilibrium.

\subsubsection{Radiative transfer}
\label{sec:radiative transfer}

The collision converts a sizable fraction of the stream kinetic energy to
internal energy. For a radiation-dominated, $\mcr\approx4$ disk with our
fiducial parameters (\cref{sec:fiducial}), the shocks dissipate energy
\numrange{\sim300}{2000} times faster than internal stresses in the disk,
corresponding to \numrange{\sim1.5}{10} times the Eddington rate
(\cref{sec:energy dissipation}). How the large amount of internal energy
created translates to emission, however, is highly uncertain.

The critical comparison here is between the inflow time and the cooling time
(\cref{sec:energy release timescales}). Since the cooling time depends on the
inner disk mass, its value can be properly determined only with \acp{MHD}
simulations that self-consistently model the amount of resupply from the stream
and the outer disk (\cref{sec:magnetic stresses}). But if the inflow time is
indeed shorter than the cooling time over a significant fraction of the mass
return time, as our hydrodynamics simulations suggest, then radiative transfer
in the inner disk is inherently time-dependent.

The relative shortness of the inflow time means radiation near the midplane is
trapped, so energy dissipation deep inside the inner disk is completely hidden
from view. The only regions that can effectively cool are those so close to the
photosphere that radiation can diffuse out before being swept into the black
hole; consequently, the outgoing luminosity is a small fraction of the total
energy dissipation rate. Even in regions from which radiation can escape,
fluctuations shorter than the local diffusion time do not imprint themselves on
the light curve.

A more careful treatment calls for three-dimensional, time-dependent radiative
transfer calculations capable of handling the high degree of asymmetry of the
system: Gas near shocks is denser and hotter than gas elsewhere, and the
outgoing material has high enough optical depth to partially obscure one side
of the inner disk.

More radiation could escape if vertical advection, for example due to magnetic
buoyancy \citep{2014ApJ...796..106J}, is important in transporting radiation
outward. Exploring this possibility requires time-dependent radiative \acp{MHD}
simulations.

The interaction among the processes above and thermalization
(\cref{sec:thermodynamics}) may depend sensitively on the specific parameters
of the system. The resulting emission may not resemble a regular \ac{TDE} or
\ac{AGN}, and may have complex temporal and spectral behavior; different
\acp{TDE} in \acp{AGN} could look entirely different.

\begin{figure}
\includegraphics{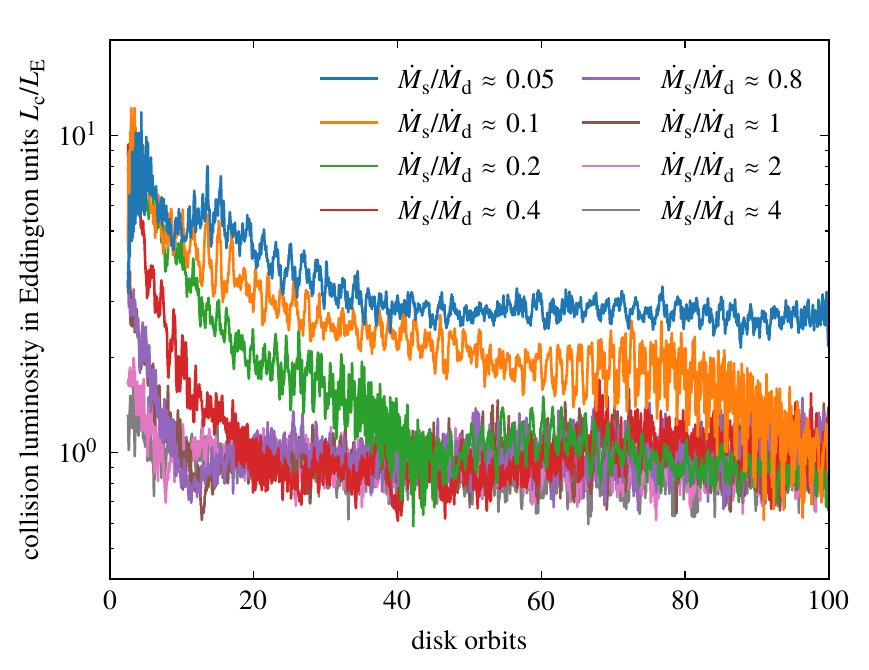}
\caption{Plot of the bolometric collision luminosity $L_\su c$ in Eddington
units as a function of time. The luminosity is estimated as $L_\su c\sim
Qt_\su{infl}/t_\su{cool}$, where $Q$ is the energy dissipation rate
(\cref{fig:energy dissipation}), $t_\su{infl}$ is the inflow time
(\cref{fig:inflow time}), and $t_\su{cool}$ is the cooling time
(\cref{sec:energy release timescales}).}
\label{fig:crude luminosity}
\end{figure}

Having laid out all these complications, we may nevertheless crudely estimate
the bolometric collision luminosity $L_\su c$ as proposed in \cref{sec:energy
release timescales}: We take the ratio $t_\su{infl}/t_\su{cool}$ of inflow to
cooling time to be roughly the fraction of energy that escapes as radiation,
and we scale the energy dissipation rate $Q$ (\cref{sec:energy dissipation}) by
it. For heavy streams, \cref{fig:crude luminosity} shows that $L_\su c$ could
be briefly very super-Eddington, then settle at a near-Eddington level. For a
radiation-dominated, $\mcr\approx4$ disk with our fiducial parameters
(\cref{sec:fiducial}), the time to steady state is \SI{\sim1}{\day}. The fact
that $L_\su c\sim L_\su E$ is not surprising in retrospect. If the inner disk
had enough time to thermalize (\cref{sec:thermodynamics}), then it would be
supported vertically against gravity by radiation pressure, and the
characteristic luminosity of such systems is $L_\su E$
\citep{2010ApJ...709..774K}.

\subsection{Further speculations}

The corona of the unperturbed inner disk may be destroyed along with the inner
disk itself (\cref{sec:overview}) if \ac{MHD} turbulence is suppressed in the
shock-driven inflow replacing it. This could explain the dip in soft
X\nobreakdash-rays lasting for a few months in the nuclear transient observed
by \citet{2019arXiv190311084T}.

The super-Eddington inflow (\cref{sec:inner disk mass}) may be connected to the
launching of a jet \citep{2011MNRAS.416.2102G, 2012ApJ...749...92K}, which has
been invoked to explain the hard X\nobreakdash-ray \citep{2011Sci...333..203B,
2011Natur.476..421B, 2012ApJ...753...77C} and radio \citep{2016Sci...351...62V}
emission of several \acp{TDE}. This possibility can be tested using future
\acp{MHD} simulations.

Acoustic waves (\cref{sec:overview}) are launched when shocks in the outer disk
push orbiting gas radially outward. Bending waves are generated when the stream
delivers misaligned angular momentum and exerts a torque on the disk. These
waves may have observable consequences on the outer disk long after the stream
has ended.

As the mass return rate dwindles toward the end of the event, the shocks would
become weaker, inner-disk gas would move inward more slowly, and the gas may
have enough time to completely cool off before reaching the black hole
(\cref{sec:energy release timescales}).

\section{Conclusions}
\label{sec:conclusions}

A small fraction of \acp{TDE} is expected to take place in \acp{AGN}. As a
first step toward understanding how \acp{TDE} in \acp{AGN} differ
observationally from \acp{TDE} in vacuum, we have conducted a suite of
simulations in which the bound debris stream of \iac{TDE} collides
perpendicularly with the pre-existing accretion disk of \iac{AGN}. Our
simulations show consistently that the collision creates shocks in the disk,
and shocks lead to extremely super-Eddington dissipation and mass inflow rates;
as a result, the disk interior to the stream impact point is heated to high
temperatures and evacuated on timescales much shorter than the mass return
time. Lighter streams merge with the disk, whereas heavier streams bore
through; in the latter case, stream gas shooting out of the other side of the
disk may interact with the disk again or with the surrounding gas.

The parameter we vary in our simulations is $\mcr$, the mass current carried by
the stream divided by the azimuthal mass current of the disk passing under the
stream footprint. It is the most important parameter governing the dynamics of
the collision because it determines the relative rates at which mass, momentum,
and kinetic energy are brought to the impact point by the stream and the disk.
As the Eddington ratio of the \ac{AGN} rises at fixed black hole mass, $\mcr$
decreases to a minimum and then increases; any given value of $\mcr$ above the
minimum can be achieved in either a weakly accreting, gas-dominated disk or a
strongly accreting, radiation-dominated disk. Typical \acp{TDE} in \acp{AGN}
have $\mcr\gtrsim1$ (\cref{sec:analytics}).

A light stream with small $\mcr$ is completely absorbed by the disk and
deflected toward the black hole (\cref{sec:outgoing material}). A heavy stream
with large $\mcr$ penetrates the disk, shooting out fluffy outgoing material on
the other side (\cref{sec:overview}). Part of the outgoing material remains
bound; it runs into the disk again eventually (\cref{sec:outgoing material}),
this time settling onto the disk instead of punching through because its
density is much lower than before (\cref{sec:overview}). Part of the outgoing
material is unbound by a kick from the disk (\cref{sec:outgoing material}); it
escapes and interacts with the surrounding gas. All these interactions are
potential energy sources for emission (\cref{sec:energetics}).

The collision dissipates mostly the kinetic energy of the stream when the
stream is light, but mostly that of the disk when the stream is heavy. The
internal energy, generated at a super-Eddington rate several orders of
magnitude above that of the unperturbed disk (\cref{sec:energy dissipation}),
raises the aspect ratio of the inner disk to order unity (\cref{sec:overview}).

The collision excites multiple shocks in the inner disk. Repeated encounters
with shocks remove angular momentum from disk gas, causing it to move speedily
inward (\cref{sec:inner disk dynamics}). Heavier streams create stronger
shocks, so much so that the inflow rate is enhanced to super-Eddington values
orders of magnitude above that of the unperturbed disk (\cref{sec:inner disk
mass}).

Rapid inflow drains the inner disk; meanwhile, the disk replenishes itself by
capturing part of the incoming stream (\cref{sec:outgoing material}). The net
effect is still a monotonic decline of the mass and surface density of the
inner disk over time, but much slower than if the stream were not captured. For
heavier streams, the surface density is lowered by an order of magnitude within
a fraction of the mass return time (\cref{sec:inner disk mass}). Resupply from
the outer disk due to magnetic stresses, not considered in our simulations,
could help the inner disk achieve mass balance within the mass return time
(\cref{sec:magnetic stresses}).

The super-Eddington inflow and energy dissipation rate does not automatically
imply super-Eddington luminosity. Because the inner disk flows inward faster
than radiation can diffuse out of it, only a small fraction of the energy
dissipated in the collision escapes as radiation, and the rest is advected
inward to the black hole (\cref{sec:energy release timescales}). Nevertheless,
the high rate at which kinetic energy is dissipated (\cref{sec:energy
dissipation}) means the bolometric luminosity may be near-Eddington
(\cref{sec:radiative transfer}). It is uncertain in which energy band the
radiation emerges; the spectrum could be far from thermal
(\cref{sec:thermodynamics}) and may not look like a standard \ac{TDE} or
\ac{AGN} at all (\cref{sec:radiative transfer}). Because the inflow time is
short and the density distribution has a complex geometry, robust observational
predictions demand more careful treatment of the three-dimensional,
time-dependent radiative transfer and the thermal evolution of the post-shock
gas (\cref{sec:radiative transfer}).

We do not know how soon the disk returns to the unperturbed state once the
stream has ended; it could take decades if inflow is controlled by \ac{MHD}
turbulence in the usual way, or a much shorter time if the cavity interior to
the impact point permits more coherent magnetic stresses to act, as often seen
in \acp{MHD} simulations (\cref{{sec:magnetic stresses}}).

We caution that, as a first step toward understanding how \acp{TDE} behave in
\acp{AGN}, our simulations considered a very restricted section of the
parameter space. The properties of the system as a function of time may depend
on such detailed parameters as the geometry and orientation of the stream
relative to the disk, the density and velocity structure of the stream, and the
properties of the unperturbed disk. Moreover, our simulations did not run long
enough to study what happens when the mass return rate has fallen
significantly, nor did they have the ability to follow the radiative and
\ac{MHD} properties of the stream and the disk. The exploration of the vast
parameter space and the wide range of physics will, no doubt, be the subject of
future work.

\ifapj\acknowledgments\else\vskip\bigskipamount\noindent\fi
The authors thank Almog Yalinewich and Nicholas Stone for insightful
discussions. C.H.C. and T.P. are supported by the ERC advanced grant
\textquote{TReX} and the CHE-ISF Center for Excellence in Astrophysics. This
work was partially supported by NSF grant AST-1715032 and by Simons Foundation
grant 559794 (J.H.K.).

\ifapj\software{Athena++ \citep{2008ApJS..178..137S}, Python, NumPy, h5py,
Matplotlib, VisIt, FFmpeg}\fi

\ifapj\bibliography{tde}\fi
\ifarxiv\printbibliography\fi
\iflocal\printbibliography\fi

\end{document}